\documentclass[review]{siamart0216}%
\usepackage{lineno}  
\usepackage[T1]{fontenc}
\usepackage[latin1]{inputenc}
\usepackage{graphicx}
\usepackage{amsmath}
\usepackage{amsfonts}
\usepackage{xcolor}
\usepackage{xspace}
\usepackage{lipsum}
\usepackage{amsfonts}
\usepackage{graphicx}
\usepackage{epstopdf}
\usepackage{algorithmic}
\providecommand{\LyX}{L\kern-.1667em\lower.25em\hbox{Y}\kern-.125emX\@}

\newcommand{\TheTitle}{Foot force models of crowd dynamics on a wobbly bridge}
\newcommand{\TheAuthors}{I. Belykh, R. Jeter, and V. Belykh}

\headers{\TheTitle}{\TheAuthors}

\title{\TheTitle\thanks{This work was supported by the National Science Foundation (USA) under Grant No. DMS-1616345  (to I.B. and R.J.) and by RSF under Grant No. 14-12-00811 and RFFI under Grant No. 15-01-08776 (to V.B.).}}

\author{Igor Belykh \thanks{Department of Mathematics and Statistics and Neuroscience Institute,
		Georgia State University, 30 Pryor Street, Atlanta, Georgia, 30303, USA.} \and Russell Jeter \footnotemark[2]\and Vladimir Belykh \thanks{Department of Mathematics, Volga State University of Water Transport, 5A, Nesterov str., Nizhny Novgorod, 603951, Russia; Department of Control Theory, Lobachevsky State University of Nizhny Novgorod, 23, Gagarin Ave., 603950, Nizhny Novgorod, Russia.} }
\begin{document}
\nolinenumbers

\maketitle

\begin{abstract}
Modern pedestrian and suspension bridges are designed using industry-standard packages, yet disastrous resonant vibrations are observed, necessitating multi-million dollar repairs. Recent examples include pedestrian induced vibrations during the openings of the Solf\'erino Bridge in Paris in 1999 and the increased bouncing of the Squibb Park Bridge in Brooklyn in 2014.  The most prominent example of an unstable lively bridge is the London Millennium Bridge which started wobbling as a result of pedestrian-bridge interactions. Pedestrian phase-locking due to footstep phase adjustment, is suspected to be the main cause of its large lateral vibrations; however, its role in the initiation of wobbling was debated. In this paper, we develop foot force models of pedestrians' response to bridge motion and detailed, yet analytically tractable, models of crowd phase-locking. We use bio-mechanically inspired models of crowd lateral movement to investigate to what degree pedestrian synchrony must be present for a bridge to wobble significantly and what is a critical crowd size. Our results can be used as a safety guideline for designing pedestrian bridges or limiting the maximum occupancy of an existing bridge. The pedestrian models can be used as ``crash test dummies'' when numerically probing a specific bridge design. This is particularly important because the US code for designing pedestrian bridges does not contain explicit guidelines that account for the collective pedestrian behavior.
\end{abstract}




\section{Introduction}
Collective behavior in mechanical systems was first discovered by Christiaan Huygens some 350 years ago \cite{huygens}. He observed two pendulum clocks, suspended on a wooden beam, spontaneously synchronize. The pendulums oscillated, remaining locked in anti-phase, while the beam remained still. The pendulums with the same support-coupling mechanism can also oscillate in-phase, in turn making the beam vibrate in anti-phase \cite{bennet,pik}. Notably, increasing the number of in-phase synchronized pendulums attached to the supporting beam leads to larger amplitudes of the swinging beam \cite{pankratova,sasha}. Originating from Huygens's experiment in simple mechanical systems, the interplay between network structure and cooperative dynamics has been extensively studied \cite{pecora,strogatz,boc,connection,motter,prx}, as cooperative dynamics and phase-locking have been shown to play an important role in the function or dysfunction of a wide spectrum of biological and technological networks \cite{gates,Winfree,newman,Motter2013}, including complex mechanical structures and pedestrian bridges \cite{focus}.

Many bridges have experienced dramatic vibrations or even have fallen down due to the effects of mechanical resonance (see \cite{failures} for a long list of bridge failures). There were two major causes for the dangerous vibrations: (i) pedestrian excitation of laterally unstable bridges  \cite{zivanovic,racic,venuti}  and (ii) wind-induced vibrations of suspension and girder bridges, including the collapse of the Tacoma Narrows Bridge  \cite{no-cause,am,green,aeroelasticity,arioli}.

Some of the most known cases of unstable pedestrian bridges are the Toda Park Bridge in Japan \cite{fujino}, Solf\'erino Bridge in Paris \cite{danbon}, the London Millennium Bridge \cite{dallard}, the Maple Valley Great Suspension Bridge in Japan \cite{nakamura-valley}, the Singapore Airport's Changi Mezzanine Bridge \cite{brownjohn}, the Clifton Suspension Bridge in Bristol, UK \cite{clifton}, and the Pedro e In\^es Footbridge in Portugal \cite{pedro}. Another a very recent example is the Squibb Park Bridge in Brooklyn \cite{brooklyn}, described  in a quote from a 2014 New York Times report \cite{times}:
``This barely two-year-old wooden structure, which cost $\$5$ million, was purposefully designed to bounce lightly with the footsteps of visitors-reminiscent of trail bridges -- but over time the movement has become more conspicuous.'' Remarkably, the bridge started to move too much, and not just up and down, but also side to side. The increased bouncing and swaying are a safety concern for pedestrians walking $50$ feet over the park.
As of February 2017, the Squibb Park Bridge remains closed to the public, and the Brooklyn Bridge Park Corporation is suing the engineering firm that designed it. A new contractor, Arup, the British engineering firm that designed the London Millennium Bridge, was recently hired to develop and oversee a plan to stabilize the Squibb Park Bridge (see the 2016 New York Times report for more details \cite{times-2}).

Interest in pedestrian collective dynamics was significantly amplified by the London Millennium Bridge which started wobbling after its opening in 2000 \cite{zivanovic,racic,venuti}.
Despite significant interest, the interaction between walking pedestrians and the London Millennium Bridge has still not been fully understood. The wobbling of the London Millennium Bridge is suspected to be initiated by a cooperative lateral excitation caused by pedestrians falling into step with the bridge's oscillations \cite{dallard,nakamura,piccardo,millennium,kuramoto,ott}.  Although, it was noted that phase-locking on the London Millennium Bridge was not perfect and pedestrians repeatedly tuned and detuned their footstep phase with the lateral bridge motion \cite{macdonald-biomechanical}.

This effect of lateral excitation of footbridges by synchronous walking was studied in recent papers \cite{dallard,nakamura,piccardo,millennium,kuramoto,ott}, including modeling the crowd synchrony by phase oscillator models \cite{millennium,kuramoto,ott}. These models explain how a relatively small synchronized crowd can initiate wobbling. As nice as these phase models are, they do not fully capture a bifurcation mechanism of the abrupt onset of wobbling oscillations that occurred when the number of pedestrians exceeded a critical value (about 165 pedestrians on the London Millennium Bridge \cite{dallard}), especially in the limiting case of identical walkers for which the phase oscillator network has no threshold for instability, and therefore the bridge would start wobbling for even a very small crowd size \cite{ott}. Rigorous analysis of the phase model system \cite{millennium,kuramoto,ott} is based on the heterogeneity of pedestrians' frequencies which is a natural assumption; however, the frequencies' density distribution is continuous, thereby implicitly assuming an infinite number of pedestrians.
Here, we develop a bio-mechanically inspired model of pedestrians' response to bridge motion and detailed, yet analytically tractable, models of crowd synchrony and phase-locking which take into account the timing and impact of pedestrian foot forces and assume a finite number of pedestrians. Our analysis predicts the critical number of pedestrians and its dependence on the frequencies of human walking and the natural frequency of the London Millennium Bridge remarkably well. Our results support the general view that pedestrian lock-in was necessary for the London Millennium Bridge to develop large-amplitude wobbling. However, our observations also suggest that the popular explanation that the wobbling of the London Millennium Bridge was initiated by crowd synchrony among pedestrians may be an oversimplification.

Surprisingly, the U.S. code and industry standard packages for designing pedestrian bridges do not explicitly address an impact of crowd collective behavior but rather relies on a higher nominal static pedestrian load and testing via the application of a periodic external force \cite{code}. The Guide Specifications for the design of pedestrian bridges  (Chapter 6)  \cite{code} limits the fundamental frequency of a bridge in the vertical and horizontal direction to 3.0 Hz and 1.3 Hz, respectively. However, these limits might not be respected, provided that there is ``phasing of loading from multiple pedestrians on the bridge at same time'' \cite{code}. The code gives no specifications on how to model and describe these phase loads. In light of this, our bio-mechanical models of pedestrian locomotion and their bi-directional interaction with a lively bridge can be incorporated into industry-standard bridge simulation packages to provide a more accurate account for the emergence of undesired crowd-induced resonant vibrations.

The layout of this paper is as follows. In \cref{section 2}, we propose pendulum models of pedestrian-bridge interaction. We first introduce an inverted pendulum model of pedestrian balance, where the control of the position of foot placement plays a significant role in lateral stabilization. We also propose a simplified, more analytically tractable model with a Van der Pol-type oscillatory mechanism of each pedestrian's gait.
In \cref{section 3}, we study bidirectional interactions between the Van der Pol-type pedestrian models and a bridge and derive explicit analytical conditions under which phase-locking and bridge wobbling appear in the system
when the crowd size exceeds a threshold value.
In \cref{section 4}, we study the pedestrian-bridge interaction model where pedestrians' lateral gaits are described by an inverted pendulum model. This analysis also reveals the threshold effect
which induces bridge wobbling at a sufficiently large crowd size. Finally, in \cref{section 5}, a discussion of the obtained results is given.

\section{Pendulum models of walker-bridge interaction}
\label{section 2}
We model lateral oscillations of the bridge and side-to-side movement of walkers by a mass-spring-damper system (Fig.~1).  The bridge is represented by a platform of mass $M$ that moves in the horizontal direction
with damping coefficient $d.$ One side of the platform is attached to the support via a spring with elasticity $k$. The platform is subjected to horizontal forces, caused by the walker response to lateral bridge motion.
The walkers are modeled by $n$ self-sustained oscillators, representing walker lateral balance and capable of adjusting their footfall forces and amplitudes in response to the lateral wobbling of the bridge. The modeling equations read:
\begin{equation}
\begin{array}{l}
\ddot{x} _i+f(x_i,\dot{x}_i)= -\ddot{y},\;\;\;\;\;
\ddot{y}+2h\dot{y}+\Omega_0^2 y =  \displaystyle{ -r\sum_{i=1}^n\ddot{x}_i, } \\
\end{array}
\label{eq:1}
\end{equation}
where the $x_i$ equation describes oscillatory side-to-side motion of the $i$-th walker. The oscillatory mechanism and response of each walker's gait to the bridge's vibrations is modeled by the function $f(x_i,\dot{x}_i),$
the effect of the bridge on $i$-th walker is described by the inertial feedback term $-\ddot{y}$. The $y$- equation describes bridge oscillations. The $i$-th walker applies sideways force $-m \ddot{x} _i$ to the bridge. Without loss of generality, each walker is assumed to have mass $m$; heterogeneity of the walkers will be introduced by non-identical gait frequencies. The ratio $r=m/(M+nm)$ represents the strength of the coupling between the walkers
and the bridge. The model (\ref{eq:1}) takes into account the role of walkers' footfall forces
and their adaptation. It extends the models \cite{millennium,kuramoto,ott} which describe the crowd synchrony where walkers are modeled by phase oscillators, and therefore only account for the timing of each walker's gait. As in \cite{millennium,kuramoto,ott}, no person-person interactions, including visual communication between the walkers and moving in dense crowds at a slower pace, are included in the model (\ref{eq:1}).

\begin{figure*}[ht]
	\begin{center}
\includegraphics[width=0.45\textwidth]{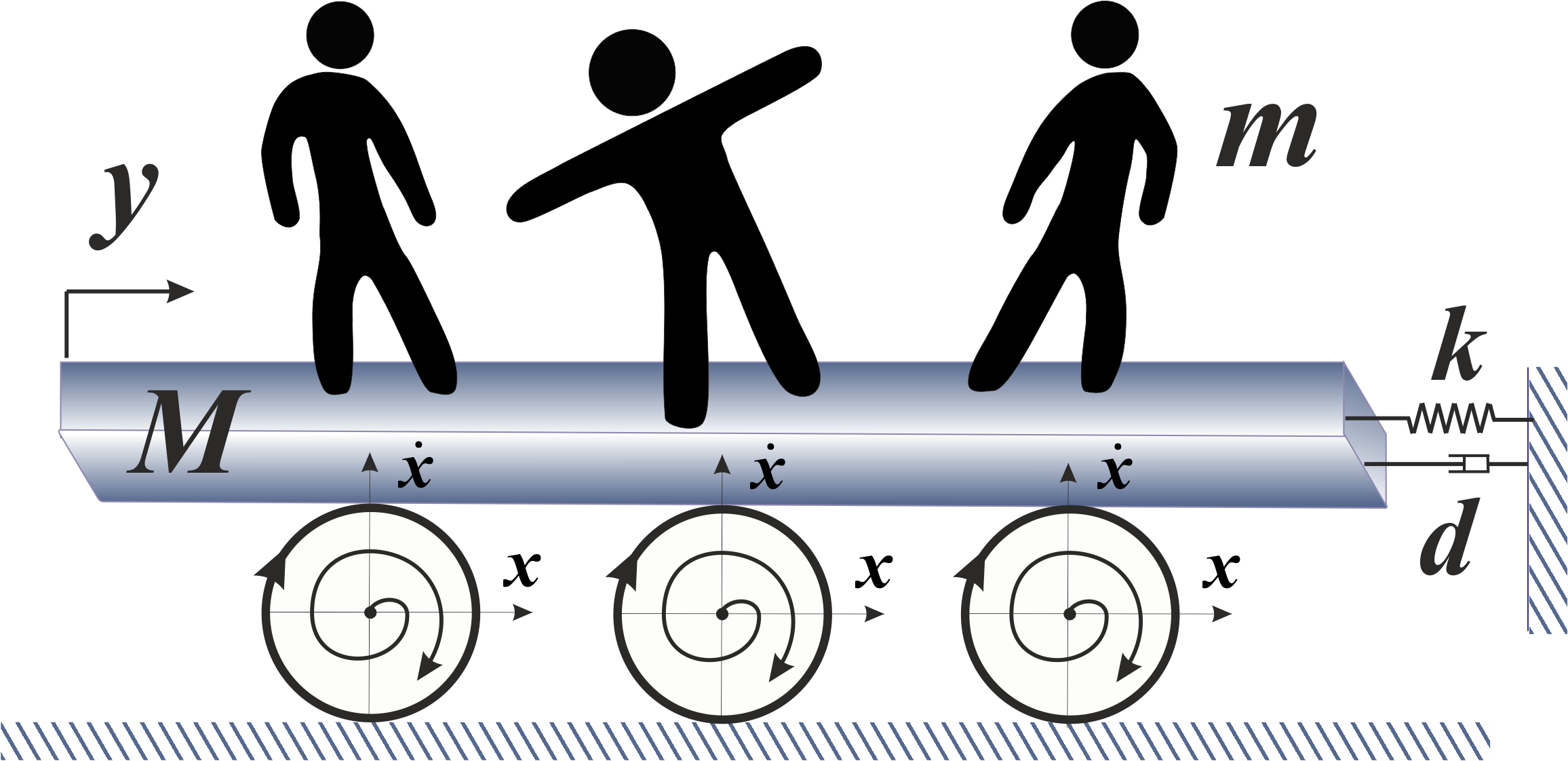} ~~~~\includegraphics[width=0.15\textwidth]{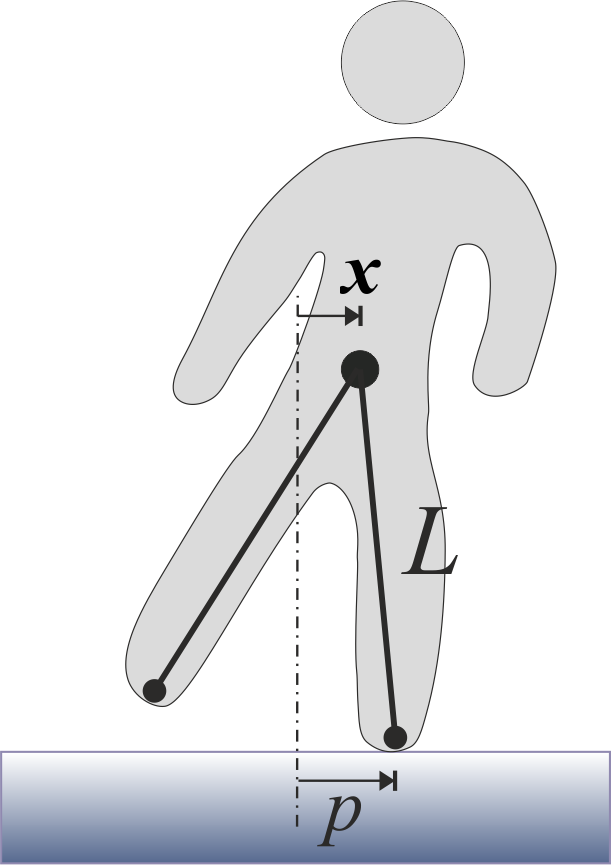}~~~~\includegraphics[width=0.27\textwidth]{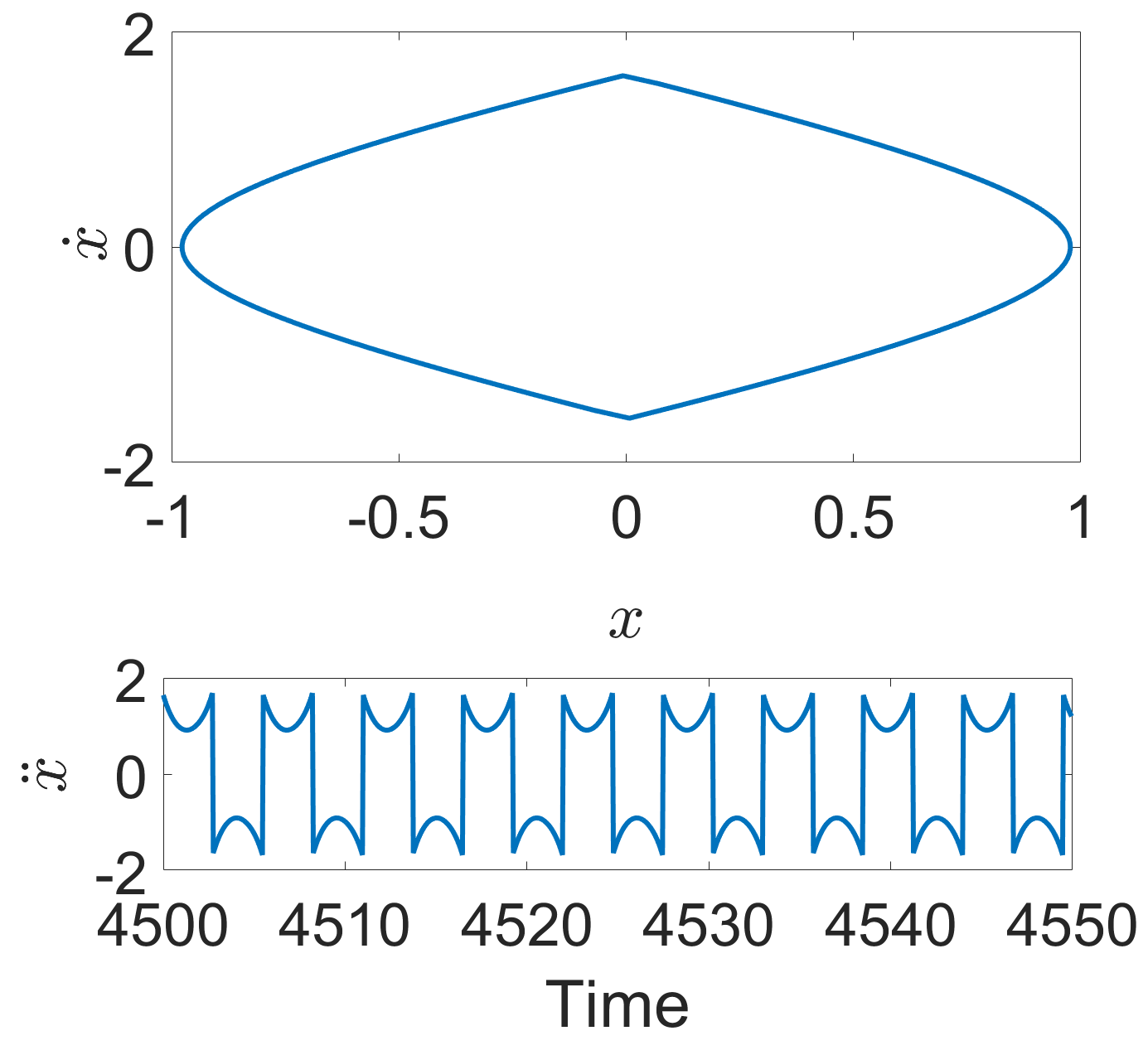}
	\end{center}
     \caption{(Left). A Huygens-type setup as a mechanical model for lateral vibrations of a bridge. The platform with mass $M$, a spring, and a damper represents lateral vibrations of the bridge $y$. Pedestrians are modeled by
     self-sustained oscillators, representing walker lateral balance, that are capable of adjusting their footfall force amplitudes $x_i$ and are subjected to lateral bridge motion. (Middle). Inverted pendulum model of pedestrian lateral movement. Variable $x$ is the lateral position of pedestrian's center of mass. Constant $p$ defines the lateral position of the center of pressure of the foot. $L$ is the inverted pendulum length. (Right). The corresponding limit cycle in the inverted pendulum model (\ref{eq:3}) along with its acceleration time series. Parameters as in Fig.~8.}
       \end{figure*}\label{fig1}

To describe $i$-th walker gait dynamics via the function $f(x_i,\dot{x}_i),$ we propose the following foot force walker models. The first bio-mechanically inspired oscillator model is an extension of the  inverted-pendulum pedestrian models \cite{barker,clifton,macdonald-inverted,macdonald-biomechanical}. The second, more-analytically tractable model uses a Van der Pol type excitation mechanism. Both models predict
the critical number of walkers ($n=165$ \cite{dallard}), beyond which the sharp onset of wobbling occurred on the London Millennium Bridge. Our study of the inverted pendulum model also indicates that crowd phase-locking stably appears in the system, and supports the observation of crowd synchrony on the London Millennium Bridge, especially during large amplitude vibrations. However, our simulations indicate that the initiation of wobbling is accompanied by multiple phase slips and nonsynchronous adjustments of pedestrians footstep such that phase lock-in mechanism might not necessarily be the main cause of initial small-amplitude wobbling.

\subsection{Inverted pendulum walker model}
Many studies in bio-mechanics have confirmed that an inverted pendulum model \cite{barker,macdonald-biomechanical} can be successfully applied to the analysis of whole body balance in the lateral direction of human walking  \cite{hof-1,hof-2,mackinnon}. The most popular bio-mechanical model \cite{macdonald-biomechanical}  of the individual walker on a rigid floor is an inverted pendulum which is comprised of a lumped mass $m$ at the center of mass supported by rigid massless legs of length $L$ (see Fig.~1 (middle)). Its equations read \cite{macdonald-biomechanical}:
\begin{equation}
\begin{array}{ll}
\ddot{x} +\omega_0^2(p-x)=0,&\;\mbox{\rm if}\;x\ge 0\;\;\mbox{\rm (right foot)}\\
\ddot{x}+\omega_0^2(-p-x)=0,&\;\mbox{\rm if}\;x< 0\;\;\mbox{\rm (left foot)},
\end{array}
\label{eq:2}
\end{equation}
where $x$ and $p$ are the horizontal displacements of the center of mass and center of pressure of the foot, respectively, and $\omega_0=\sqrt{g/L}$ with $g$ being the acceleration due to gravity and $L$ being the distance from the center of mass to the center of pressure. While this model captures the main time characteristics of human walking such as lateral displacement, velocity, and acceleration rather precisely (see \cite{macdonald-inverted,macdonald-biomechanical} for the details), it has several limitations, especially when used in the coupled walker-bridge model (\ref{eq:1}). From a dynamical systems theory perspective, this is a conservative system with no damping; its phase portrait is formed by three fixed points: a center at the origin which is confined in a diamond-shape domain, bounded by the separatrices of two saddles at $(p,0)$ and $(-p,0).$ The periodic motion is then governed by closed conservative curves of the center fixed point. As a result, the amplitude of a walker's lateral gait depends on the initial conditions (via the corresponding level of the closed curves of the center) and is determined by the size of the first step, even though this step might be too big or too small.
Note $-x$, not $+x$ in the equations (\ref{eq:2}). This implies that the solutions of each of the two systems for the right and left feet, comprising the piecewise linear system (\ref{eq:2}), are not cosine and sine functions as with the harmonic oscillator but instead hyperbolic functions. Therefore, the closed curves, ``glued'' from the solutions of the two systems, have a distinct diamond-like shape, yielding realistic time series for lateral velocity and acceleration of human walking (see Fig.~1 (right)).

We propose the following modification of the pedestrian model (\ref{eq:2}) to make it a self-sustained oscillator:
\begin{equation}
\begin{array}{l}
\ddot{x} =\omega_0^2(x-p)-\lambda(\dot{x}^2-\nu^2(x-p)^2+\nu^2a^2)\dot{x},\; \mbox{\rm if}\;x\ge 0 \; \mbox{\rm (right foot)},\\
\ddot{x} =\omega_0^2(x+p)-\lambda(\dot{x}^2-\nu^2(x+p)^2+\nu^2a^2)\dot{x},\;\mbox{\rm if}\;x< 0 \; \mbox{\rm (left foot)},
\end{array}
\label{eq:3}
\end{equation}
where $\lambda$ is a damping parameter, and $a$ is a parameter that controls the amplitude of the limit cycle (lateral foot displacement). Parameter $\nu$ controls the period of the limit cycle. Note that system (\ref{eq:3}) has a stable limit cycle (Fig.~1 (right)) which coincides with a level of the nonlinear center of conservative system (\ref{eq:2}) for $\nu=\omega_0$. The period of the limit cycle can be calculated explicitly as the limit cycle is composed of two glued solutions for the right and left foot systems where $x=a\cosh \nu t$ and $\dot{x}=a \nu \sinh \nu t$ (for $\nu=\omega_0$) correspond to one side of the combined limit cycle, glued at $x=0$. The use of system (\ref{eq:3}) instead of conservative system (\ref{eq:2}) in the coupled system allows one to analyze the collective behavior of the walkers-bridge interaction much more effectively, yet preserving the main realistic features of human walking.  System (\ref{eq:3}) can also be written in a more convenient form:
\begin{equation}
\begin{array}{l}
\ddot{x} +\lambda(\dot{x}^2-\nu^2 x^2+\nu^2a^2)\dot{x}-\omega_0^2 x=0,\;\;\;x+p({\rm mod}\; 2p),\\
\end{array}
\label{eq:4}
\end{equation}
defined on a cylinder and obtained by shifting $x$ to $x+p$. As a result, the two saddles shift to $x=0$ and $x=2p$, and the limit cycle becomes centered around $x=p.$ In the following, we will use (\ref{eq:3}) as the main model of the individual walker in our numerical studies of the coupled walker-bridge model (\ref{eq:1}) with $f(x_i,\dot{x}_i),$ defined via the RHS of model (\ref{eq:4}). The analytical study of the coupled model (\ref{eq:1})-(\ref{eq:4}) is somewhat cumbersome and will be reported in a more technical publication; we will further simplify the walker-bridge model (\ref{eq:1})-(\ref{eq:4}), incorporating a Van der Pol-type excitation mechanism and making the analysis more elegant and manageable.

\subsection{Van der Pol-type pendulum model}
An oscillatory mechanism of each walker's gait, modeled by the function $f(x_i,\dot{x}_i)=\lambda(\dot{x}_i^2-\nu^2 x_i^2+\nu^2a^2)\dot{x}_i-\omega_0^2 x_i$ in (\ref{eq:4}),  can be similarly
described by the Van der Pol-type term $f(x_i,\dot{x}_i)=\lambda_i (\dot{x}_i^2+x_i^2-a^2)\dot{x}_i +\omega_i ^2x_i.$ Therefore, the walker-bridge model (\ref{eq:1})-(\ref{eq:4}) can be transformed into the following system
\begin{equation}
\begin{array}{l}
\ddot{x} _i+\lambda_i (\dot{x}_i^2+x_i^2-a^2)\dot{x}_i +\omega_i ^2x_i= -\ddot{y},\\
\ddot{y}+2h\dot{y}+\Omega^2 y =  \displaystyle{ -r\sum_{i=1}^n\ddot{x}_i, } \\
\end{array}
\label{eq:b1}
\end{equation}
where the dimensionless time $\tau=\nu t,$  $\omega_i=\omega_0^{(i)}/\nu,$ and $\Omega=\Omega_0/\nu.$ Walkers are assumed to have different natural frequencies $\omega_i,$ randomly distributed within the range $[\omega_{-},\omega_{+}].$
The $i$-th Van der Pol-type oscillator has a limit cycle with a frequency that lies between $\omega_i$ and $1.$ Note that the equations for the limit cycle of the individual Van der Pol-type oscillator cannot be found explicitly, yet we will be able to find phase-locked solutions in the coupled pedestrian-bridge model (\ref{eq:b1}).

It is important to emphasize that this model is an extension of the London Millennium Bridge reduced (phase) models \cite{millennium,kuramoto,ott} where the dynamics of each walker is modeled by a simple phase oscillator with phase $\Theta_i.$ In terms of the model (\ref{eq:b1}), the Eckhard  {\it et al.} model  \cite{kuramoto}
reads $\dot{\Theta}_i=\omega_i-c_i \ddot{y} \cos(\Theta_i+\xi_i);\; \ddot{y}+2h\dot{y}+\Omega_0^2 y =  \displaystyle{ c\sum_{i=1}^n \cos\Theta_i},$  the description of the parameters can be found in \cite{kuramoto}.
The frequencies $\omega_i$ are randomly chosen from a continuous density distribution $P(\omega),$ thereby assuming an infinite number of pedestrians.  The Eckhard  {\it et al.} model allows for a nice reduction to a Kuramoto-type network and derivation of analytical estimates on the degree of coherence among the phase oscillators that causes the bridge oscillations to amplify. That is, the wobbling begins as a Kuramoto-type order parameter gradually increases in time when the size of the synchronized group increases in time \cite{millennium,kuramoto,ott}. However, these phase models, based on the widespread notion of footstep phase adjustment as the main mechanism of synchronous lateral excitation, have the above mentioned limitations, including the absence of a threshold on the critical number of {\it identical} pedestrians, initiating abrupt wobbling. The presence of the feedback term $\ddot{y}$ in the $x_i$-equation of the model (\ref{eq:b1}) has been the main obstacle, preventing the rigorous transformation of (\ref{eq:b1}) to an analytically tractable model in polar coordinates. As these limitations call for more detailed models, also taking into account foot forces and position of foot placement, we use the bio-mechanically inspired model (\ref{eq:1})-(\ref{eq:4}) and its Van der Pol-type simplification (\ref{eq:1})-(\ref{eq:b1}) to make progress towards mathematical understanding of the mechanism of the London Millennium Bridge's vibrations. We start off  with rigorous analysis of the model (\ref{eq:1})-(\ref{eq:b1}) and its implications to the London Millennium Bridge. We will derive
an explicit bound on the critical crowd size needed for the onset of wobbling in the case of both identical and non-identical pedestrians. We will also numerically validate this bound for both inverted pendulum and Van der Pol-type models and obtain an excellent fit.

\section{Predicting the onset of bridge wobbling}
\label{section 3}
It is important to emphasize that such a high-dimensional, nonlinear pedestrian-bridge model (\ref{eq:b1}) is not expected to be analytically tractable in general.
As a result, finding its close-form analytical solution which corresponds to phase-locked behavior at an unknown common frequency seems to be out of reach for the existing methods. Pedestrian-bridge interactions in the model (\ref{eq:b1}) are strong and the model contains no small parameters. Therefore, the averaging methods which are typically used in the Kuramoto-type phase models \cite{millennium,kuramoto,ott} or phase models of Josephson junctions with a common load \cite{kurt} to identify the common frequency of phase-locking cannot be applied. Instead, we will solve the inverse problem and synthesize an analytical solution which generates phase-locking at a desired, pre-defined frequency. Towards this goal, we will derive explicit conditions on the intrinsic parameters of pedestrians, the bridge, and the crowd size under which this phase-locked solution appears in the system (\ref{eq:b1}) when the crowd size exceeds a threshold value. This periodic solution is induced by the adjustment of pedestrians' frequencies to the frequency of the bridge, yielding the one-to-one phase locking frequency ratio between the pedestrians and bridge.

We seek the periodic solutions in the form
\begin{equation}
\begin{array}{l}
x_i(t)=B_i\sin(t+\varphi_i)$ and $y(t)=A\sin(t+\psi),
\end{array}
\label{eq:sol1}
\end{equation}
 where $B_i$ and $\varphi_i$ are the amplitude and phase of the $i$-th oscillator (pedestrian) and $A$ and $\psi$ are related to the bridge.  Therefore, we set the frequency of the pedestrian-bridge phase-locked motion  equal to one. The harmonic solutions $x_i$ and $y$ with frequency one do not exist in the pedestrian-bridge model (\ref{eq:b1}) with arbitrarily chosen natural frequencies of the pedestrians and the bridge. However, we will
 give an explicit recipe of how to choose the natural frequencies and other intrinsic parameters  which yield and preserve the harmonic solutions $x_i$ and $y$ with frequency one when the crowd size increases.
 Therefore, we will be able to conduct the assumed harmonic solutions through the ``eye of the needle'' of the otherwise analytically intractable model (\ref{eq:b1}) and demonstrate that these phase-locked solutions appear when the number of pedestrians exceed the threshold value, causing an abrupt onset of bridge wobbling.

\subsection{Identical Van der Pol-type walkers}
\subsubsection{Analytical predictions}
We begin with the worst-case scenario for the onset of wobbling, caused by identical pedestrians with uniform $\omega_i=\omega,$ when, for example, British Royal Guards nearly equal in size and weight walk across the bridge, breaking step. This case is out of reach for the phase oscillator models that would become phase-locked at an infinitesimal crowd size  \cite{millennium,kuramoto,ott}, similar to the Kuramoto model with uniform $\omega$ that synchronizes at infinitesimally small coupling values \cite{acebron}.

The phase-locked solutions (\ref{eq:sol1}) in the pedestrian-bridge model (\ref{eq:b1}) with identical oscillators correspond to complete synchronization among the pedestrians such that
$x_1=...=x_n=x(t)=B \sin(t+\varphi)$ and $y(t)=A\sin(t+\psi)$ as $B_i=B$ and $\varphi_i=\varphi,$ $i=1,...,n.$ Substituting the solutions for $x$ and $y$ along with their derivatives into the first equation of system (\ref{eq:b1}) yields
\begin{equation}
\begin{array}{l}
B(\omega^2-1) \sin (t+\varphi) +\lambda B(B^2-a^2) \cos(t+\varphi)=A \sin (t+\psi).
\end{array}
\label{eq:i1}
\end{equation}
Introducing the angle $\beta$ such that
\begin{equation}
\begin{array}{l}
\cos \beta=\frac{\omega^2-1}{\sqrt{(\omega^2-1)^2+\lambda^2(B^2-a^2)^2}},\;\;
\sin \beta=\frac{\lambda (B^2-a^2)}{\sqrt{(\omega^2-1)^2+\lambda^2(B^2-a^2)^2}},
\end{array}
\label{eq:i2}
\end{equation}
we obtain
\begin{equation}
\begin{array}{l}
B \sqrt{(\omega^2-1)^2+\lambda^2(B^2-a^2)^2} [\cos \beta \sin (t+\varphi) +\sin \beta \cos(t+\varphi)]=A\sin (t+\psi).
\end{array}
\label{eq:i3}
\end{equation}
Using a trigonometric identity $\cos \beta \sin (t+\varphi) +\sin \beta \cos(t+\varphi)=\sin(t+\varphi+\beta),$ we turn the equation (\ref{eq:i3}) into
\[
\begin{array}{l}
B \sqrt{(\omega^2-1)^2+\lambda^2(B^2-a^2)^2} \sin(t+\varphi+\beta) =A \sin (t+\psi).
\end{array}
\]
This equation holds true if
\[
\begin{array}{l}
A=B \sqrt{(\omega^2-1)^2+\lambda^2(B^2-a^2)^2},\;\mbox{\rm and}\;\beta=\psi-\varphi.
\end{array}
\label{eq:i5}
\]
Taking into account formulas (\ref{eq:i2}),  we obtain the following equations for amplitude and phase balances between the pedestrians and the bridge
\begin{equation}
A=\sqrt{B^2(\omega^2-1)^2+\lambda^2(B^2-a^2)^2B^2},
\label{eq:i6}
\end{equation}
\begin{equation}
\tan(\varphi-\psi)=-\lambda\frac{B^2-a^2}{\omega^2-1}.
\label{eq:i7}
\end{equation}
Clearly, we need extra balance equation to identify the unknown parameters $A, B$, $\varphi, \psi$. Similarly, substituting the solutions into the second equation of (\ref{eq:b1}), we obtain
\begin{equation}
\begin{array}{l}
A=\frac{rn}{\sqrt{\Delta}}B,
\end{array}
\label{eq:b3}
\end{equation}
\begin{equation}
\begin{array}{l}
\tan(\varphi-\psi)=\frac{2h}{1-\Omega^2},
\end{array}
\label{eq:i8}
\end{equation}
where $\Delta=(\Omega^2-1)^2+4h^2$.

Comparing the amplitude balance equations (\ref{eq:i6}) and (\ref{eq:b3}) yields
\begin{equation}
\begin{array}{l}
B^2=a^2+\frac{1}{\lambda}\sqrt{\frac{(rn)^2}{\Delta}-(\omega^2-1)^2}.
\end{array}
\label{eq:i9}
\end{equation}
Therefore, a necessary condition for the existence of the phase-locked solutions $x(t)$ and $y(t),$ allowing the $x$-amplitude to be a real value, is $\frac{(rn)^2}{\Delta} \ge (\omega^2-1)^2.$ This condition gives the following bound on the pedestrians $n>n_c,$ required for the phase-locked solution to exist, where exceeding the critical crowd size $n_c$ leads to the abrupt onset of bridge wobbling:
\begin{equation}
\frac{m n_c}{m n_c+M}=|\omega^2-1|\sqrt{(\Omega^2-1)^2+4h^2}.
\label{eq:i10}
\end{equation}

Figure~2 illustrates this condition and presents mutual arrangements of the amplitude balance curves (\ref{eq:i6}) and (\ref{eq:b3}) as a function of the crowd size $n.$
Intersections between the cubic-like curve $A=\sqrt{B^2(\omega^2-1)^2+\lambda^2(B^2-a^2)^2B^2}$ (blue line) and a straight line $A=\frac{rn}{\sqrt{\Delta}}B$ (dashed black) correspond to the existence of the phase-locked solutions (\ref{eq:sol1}). The parameters are chosen to make $n_c=165$ \cite{dallard}, beyond which the sharp onset of wobbling occurred on the London Millennium Bridge.
For $n<165,$ the slope of the straight line is not large enough to ensure the intersection with the cubic-like curve, whereas $n_c=165$ yields the emergence of the phase-locked solution (\ref{eq:sol1}) and initiates bridge wobbling.

\begin{figure*}[ht]
	\begin{center}
\includegraphics[width=0.6\textwidth]{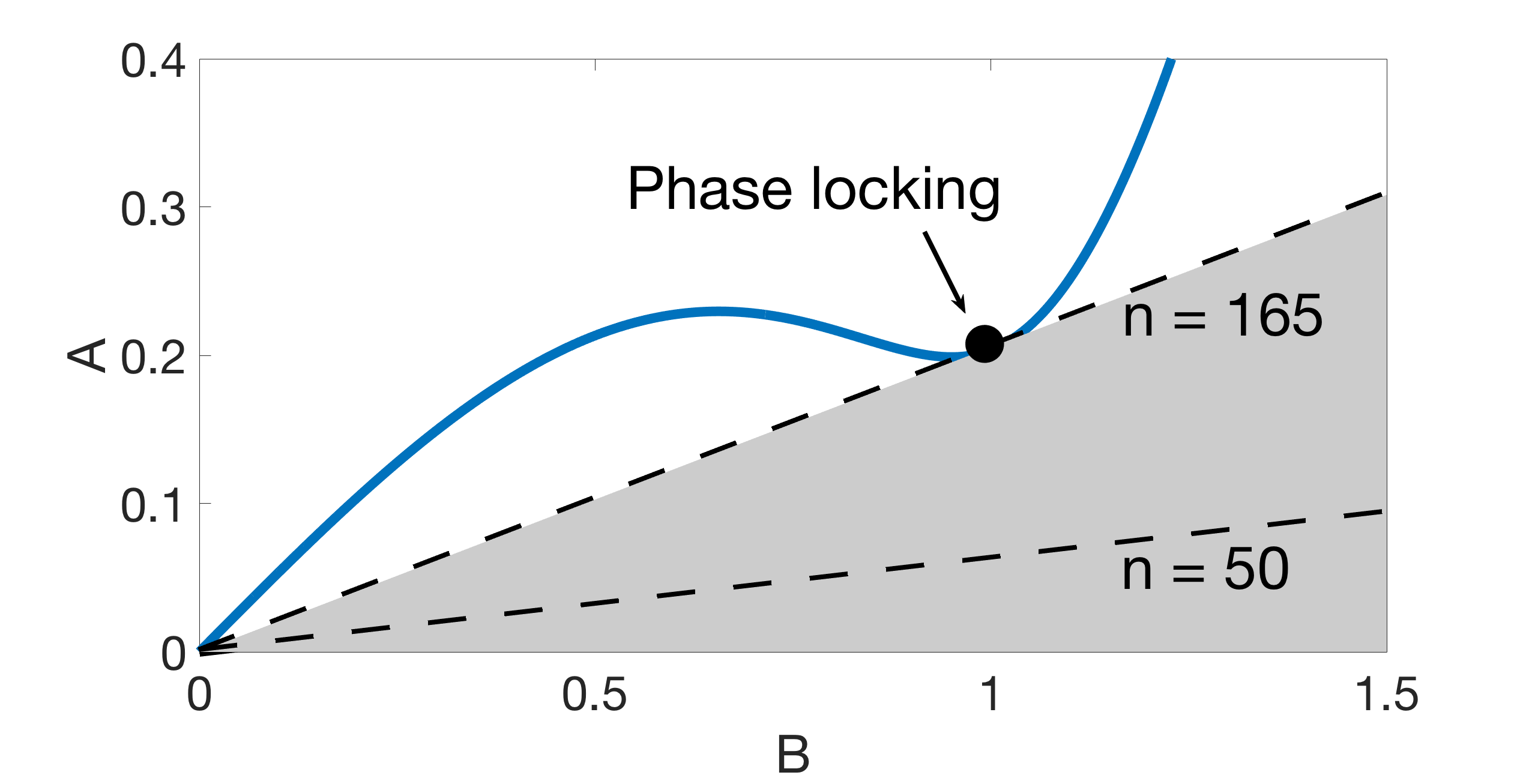}
\end{center}
     \caption{Illustration of the amplitude balance conditions (\ref{eq:i6}) and (\ref{eq:b3}). The light gray area: the absence of wobbling, guaranteed by (\ref{eq:i10}).
     Intersections between the blue (solid) curve (\ref{eq:i6}) with a dashed line (\ref{eq:b3}) generate phase-locked solutions for the number of pedestrians exceeding a critical number $n_c=165$.
     Parameters as in Fig.~3. The solid circle relates to the corresponding point in Fig.~3.}
     \end{figure*}

However, for the phase-locked solution (\ref{eq:sol1}) with frequency one to exist for $n \ge n_c,$ an additional constraint such as the phase balance must be imposed.
Comparing the phase balance equations (\ref{eq:i7}) and (\ref{eq:i8}), we obtain
\begin{equation}
\begin{array}{l}
B^2=a^2+\frac{2h(\omega^2-1)}{\lambda(\Omega^2-1)},
\end{array}
\label{eq:i11}
\end{equation}
Equating the right-hand sides (RHS) of (\ref{eq:i8}) and (\ref{eq:i11}), we derive the following expression which links the parameters of the pedestrians and the bridge and indicates, in particular, how the natural
frequency $\omega$ of the pedestrian should be chosen to satisfy the amplitude-phase balances (\ref{eq:i8})-(\ref{eq:i11}):
\begin{equation}
\begin{array}{l}
\omega^2=1+\frac{mn}{C(mn+M),}\;\mbox{\rm where}\;\;C=\Omega^2-1+\frac{4h^2}{\Omega^2-1}.
\end{array}
\label{eq:i12}
\end{equation}
\begin{figure*}[ht]
	\begin{center}
\includegraphics[width=0.65\textwidth]{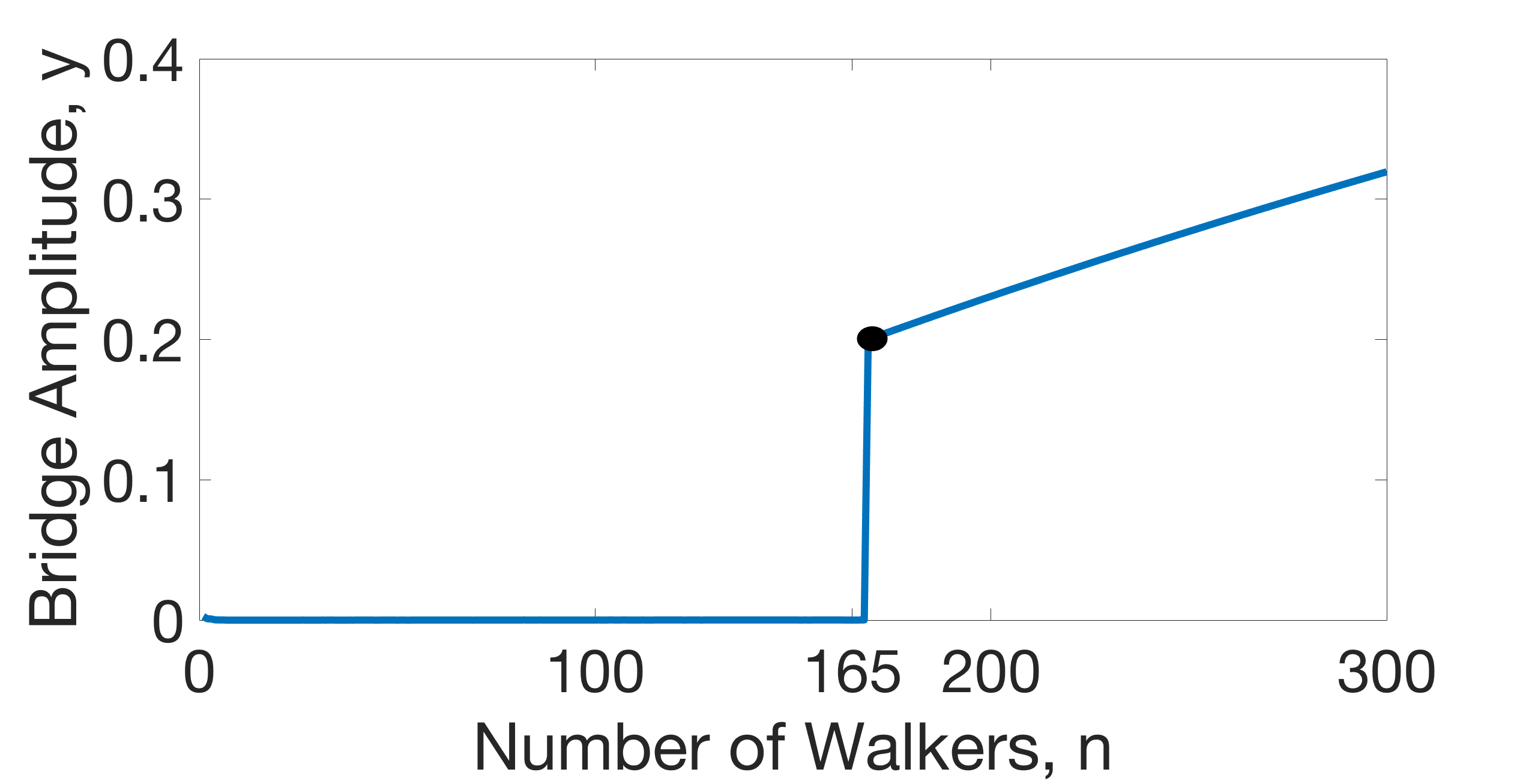}\\
\includegraphics[width=0.65\textwidth]{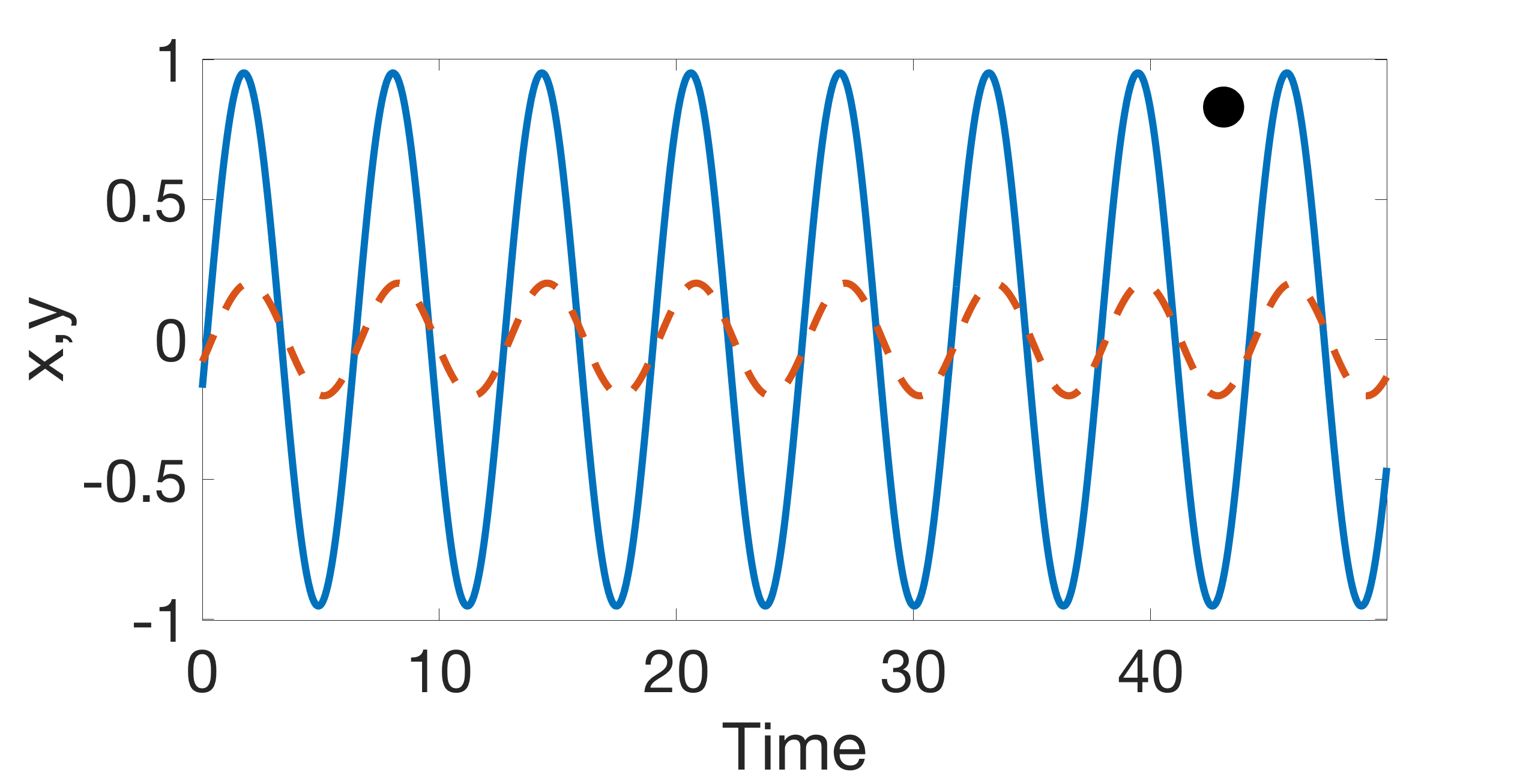}\\
\includegraphics[width=0.65\textwidth]{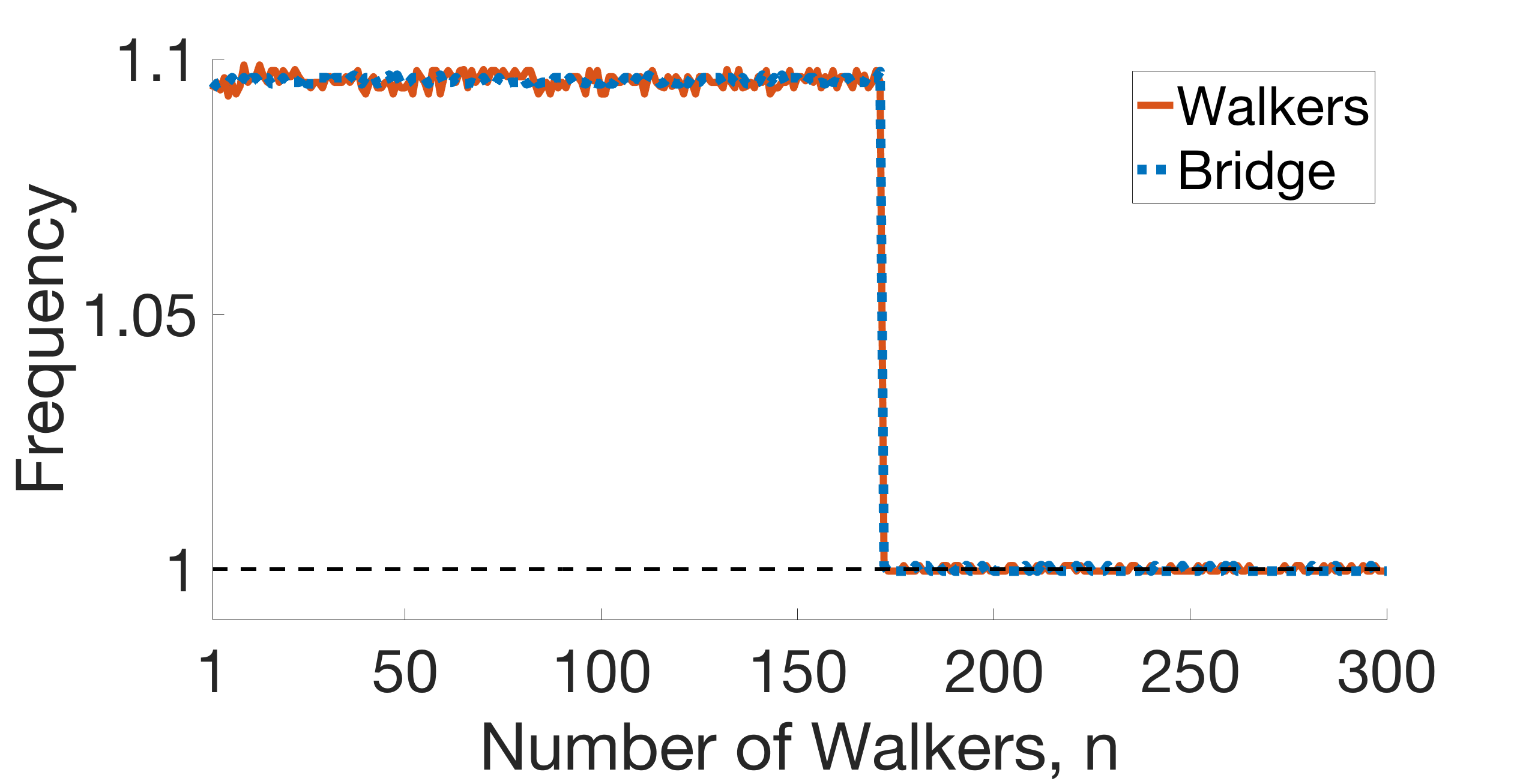}
\end{center}
     \caption{Numerical verification of the analytical prediction (\ref{eq:i10})-(\ref{eq:i12}): phase locking with frequency one.
             (Top): Abrupt onset of bridge wobbling from random initial conditions as a function of the number of walkers in the Van der Pol-type pedestrian-bridge model (\ref{eq:b1}). (Middle): Time-series of phase-locking among the pedestrians (blue solid line) and with the bridge (red dashed line) at the common frequency equal to one. The solid circles correspond to $n=165$ and also relate to the point in Fig.~2. (Bottom): Adjustment of the averaged pedestrian and bridge frequencies. Note the bridge frequency jitters due to bridge beating, prior to the onset of phase locking. The averaged bridge frequency is calculated via averaging the distances between subsequent peaks of quasi-periodic beating induced by non-synchronized pedestrians' motions. The emergence of phase-locking at frequency one at $n=165$ is in perfect agreement with the analytical prediction. Parameters are $\Omega = 1.2$, $\lambda = 0.5$, $h = 0.05$, $m = 70$, $M = 113,000$, $a = 1.0$. For $n\in [1,165],$ the natural pedestrian frequency $\omega=1.097$ is fixed and chosen via (\ref{eq:i12}) to ensure that the phase-locked solution  appears at $n_c=165.$ For $n\in [166,300],$ the natural frequency $\omega$ is varied via (\ref{eq:i12}) to preserve the amplitude-phase balance.}
     \end{figure*}

\subsubsection{Numerical verification: phase-locking at frequency one}
To test our analytical prediction (\ref{eq:i10})-(\ref{eq:i12}), we have performed a series of numerical simulations (see Fig.~3) which reveal a sharp onset of bridge wobbling when the crowd size exceeds its critical value $n_c=165.$ This onset is induced by the emergence of phase-locking among the pedestrians and the bridge at the pre-defined frequency one. For this solution to exist for $n \ge n_c=165,$ we calculate the natural frequency $\omega$ via the amplitude-balance condition  (\ref{eq:i12}). For the range of crowd sizes $n\in [1,165]$ which involves the abrupt transition from the absence of wobbling to significant wobbling, we choose and fix $\omega=1.097$ via (\ref{eq:i12}) using $n=165$ and the parameters given in Fig.~3. This guarantees that the desired phase-locked solution with frequency one does not exist for $n<165$ and emerges at $n=165.$ To preserve the existence of phase-locked solution for large crowd sizes $n>165,$ we vary $\omega$ via (\ref{eq:i12}) for each $n>165$ to satisfy the balance condition (\ref{eq:i12}). This makes $\omega$ increase from $1.098$ to $1.15$ when $n$ increases from $166$ to $300.$ Note that the analytical conditions (\ref{eq:i10})-(\ref{eq:i12}) only guarantee the existence or absence of the phase-locked solution for a given $n$ and do not account for its stability.
The pedestrian-bridge model (\ref{eq:b1}) contains a free parameter $\lambda$ which is not explicitly present in the existence conditions (\ref{eq:i10})-(\ref{eq:i12}) so that it can be tuned to make the desired phase-locked solution to become stable. Remarkably, the phase solution with frequency one becomes stable as long as it emerges in a wide range of parameter $\lambda.$ Therefore, we made little effort to tune $\lambda$ to $0.5.$
Our multiple repetitive simulations from random initial conditions with $x_i(0),$ $i=1,...,n$ chosen from the range $[-1,1]$ have indicated the emergence of the same stable phase-locked rhythm, suggesting that the phase-locked solution is globally stable.

\subsubsection{Phase-locking at a different frequency}
\begin{figure}
	\begin{center}
\includegraphics[width=0.65\textwidth]{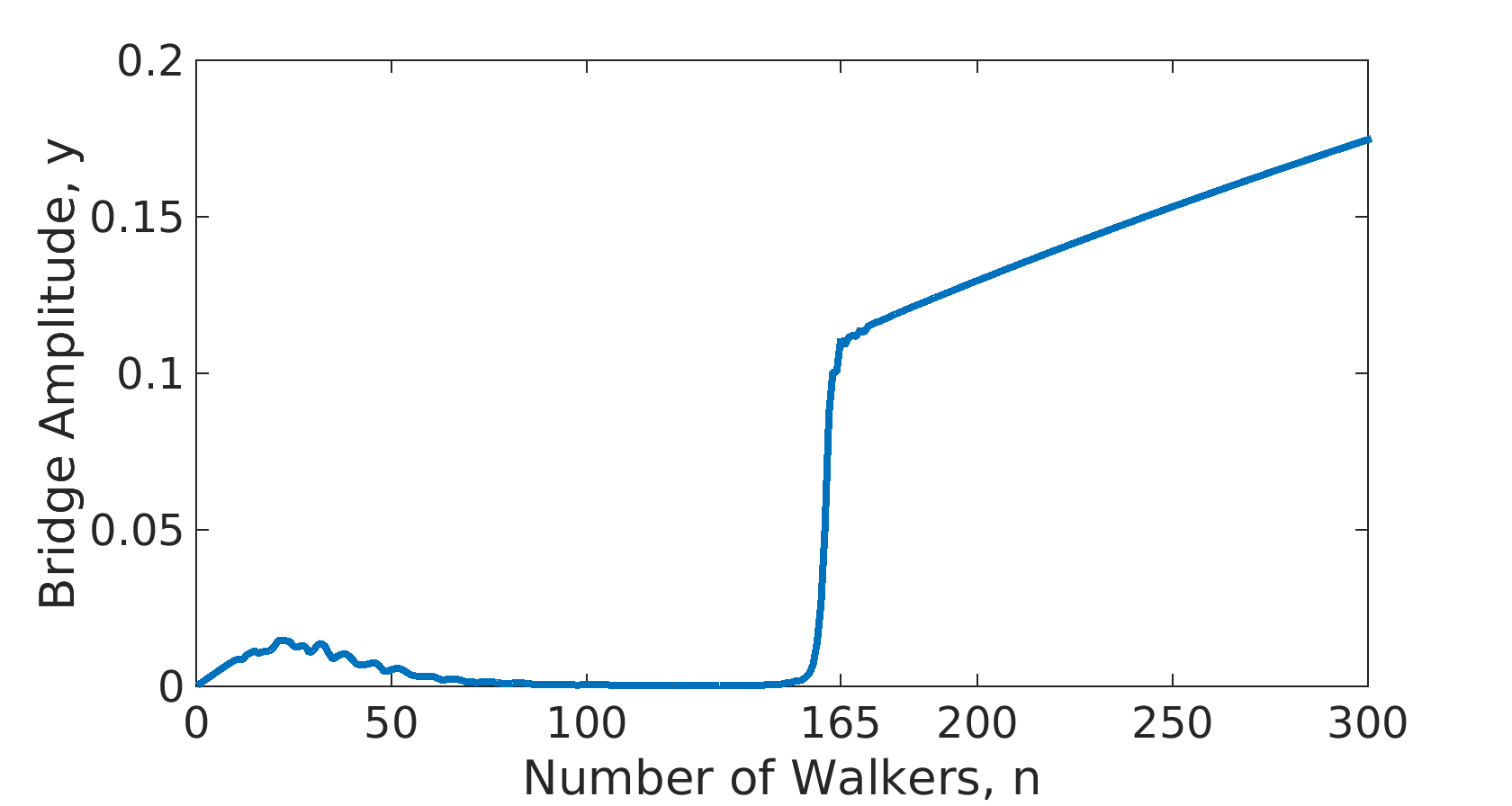}
 \includegraphics[width=0.65\textwidth]{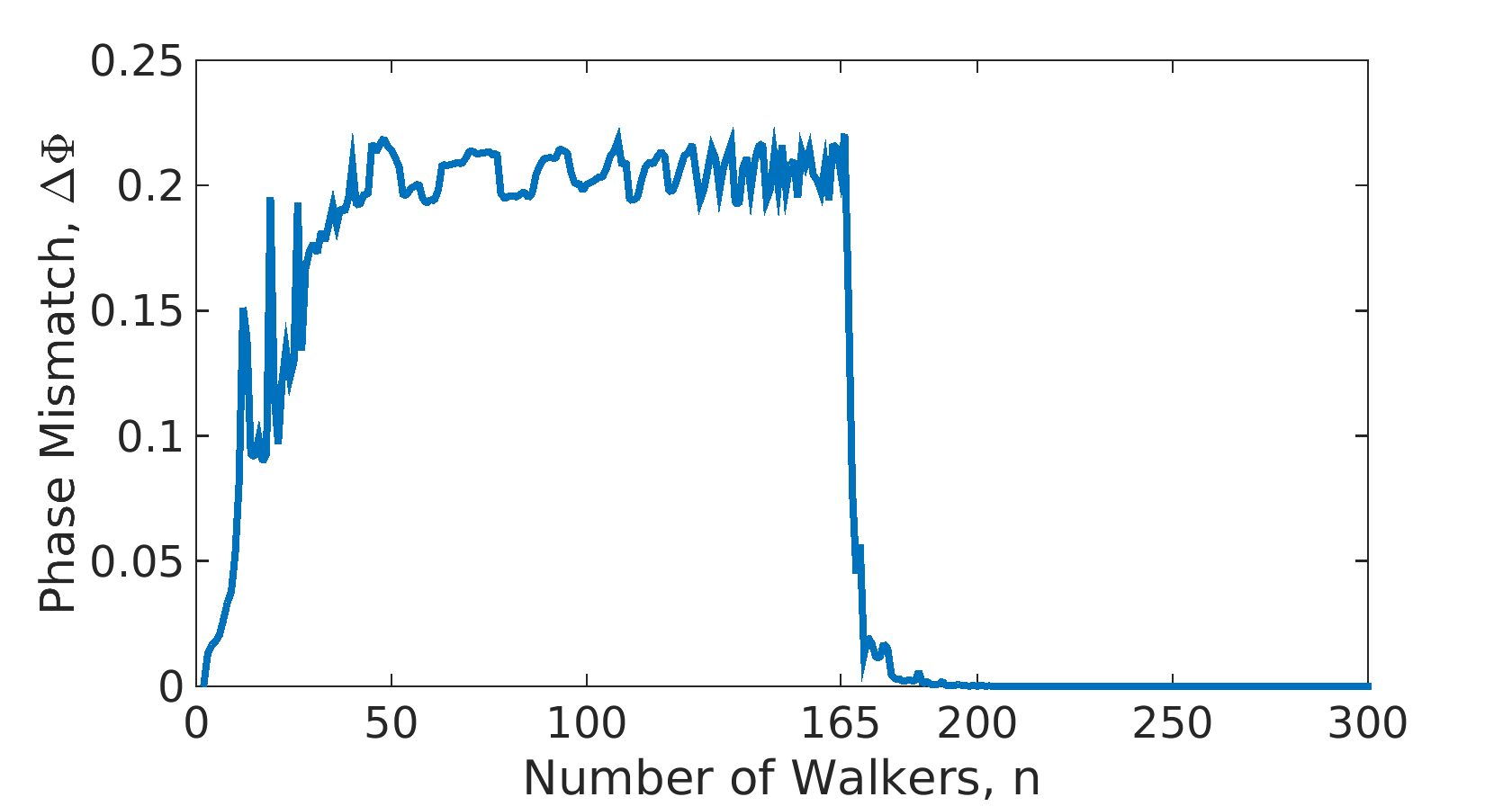} \includegraphics[width=0.65\textwidth]{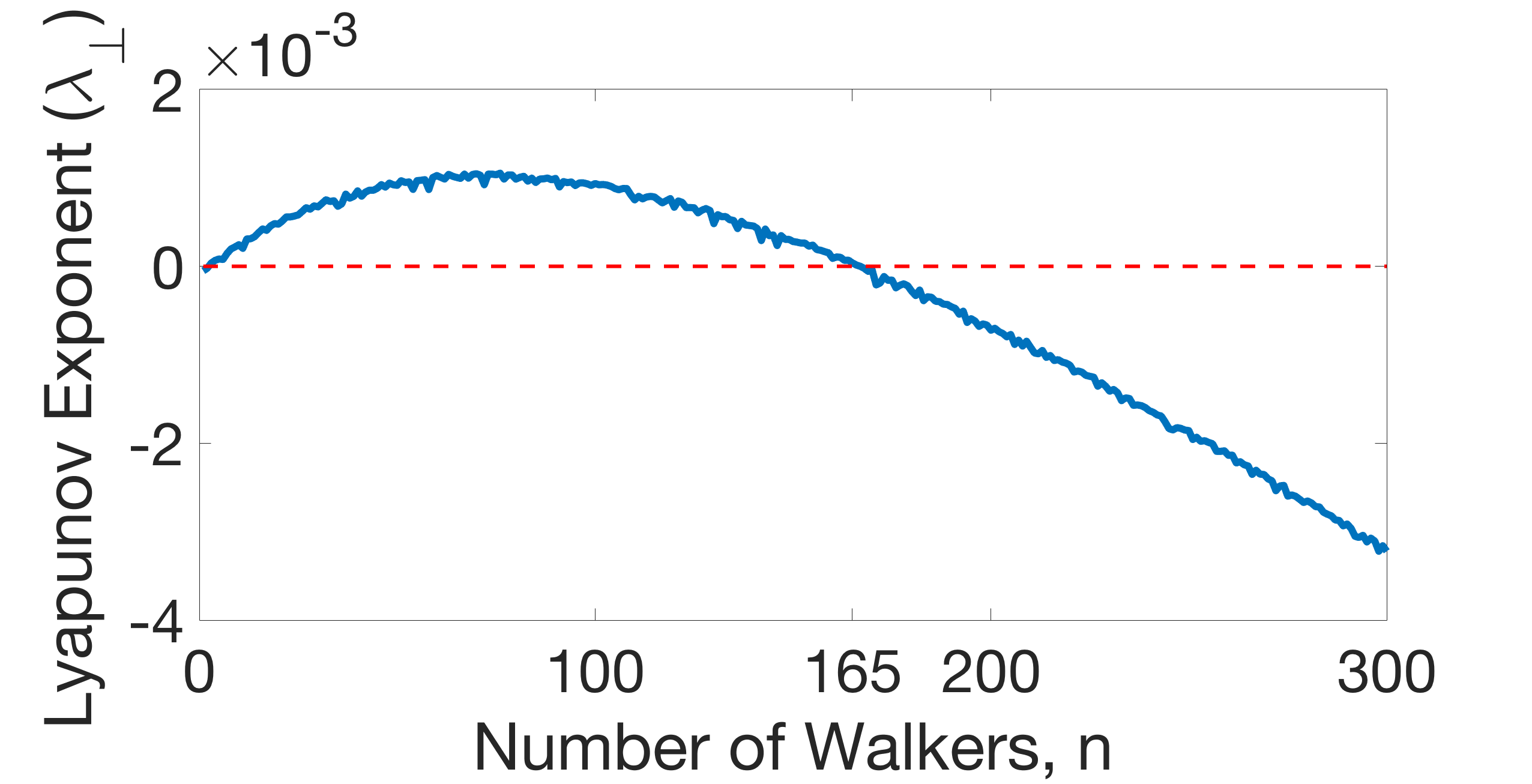}
\end{center}
     \caption{Analytically intractable case of $\Omega=1.$  Identical pedestrians with $\omega=0.73,$ $\lambda=0.23,$ $a=1.$ Other parameters are chosen to fit the data for the London Millennium Bridge: $h=0.05,$ $M=113,000,$ and $m=70$. The onset of bridge wobbling (top) coincides with pedestrian phase-locking ($\Delta \Phi = 0$) at $n=165$ (middle). Notice the appearance and disappearance of the small bump in bridge wobbling $n\in[10,40]$ due to a chimera state where a subgroup of pedestrians becomes phase-locked [not shown]. Crowd phase-locking measured by the average phase difference $\Delta \Phi$ of pedestrians.
     (Bottom). The largest transversal Lyapunov exponent $\lambda_\perp$ for the stability of complete phase-locking between pedestrians.  Its negative values (occurring around $n \approx 165$) indicate the stability of the phase-locked solution.}
     \label{fig3}
       \end{figure}
\begin{figure}
	\begin{center}
\includegraphics[width=0.49\textwidth]{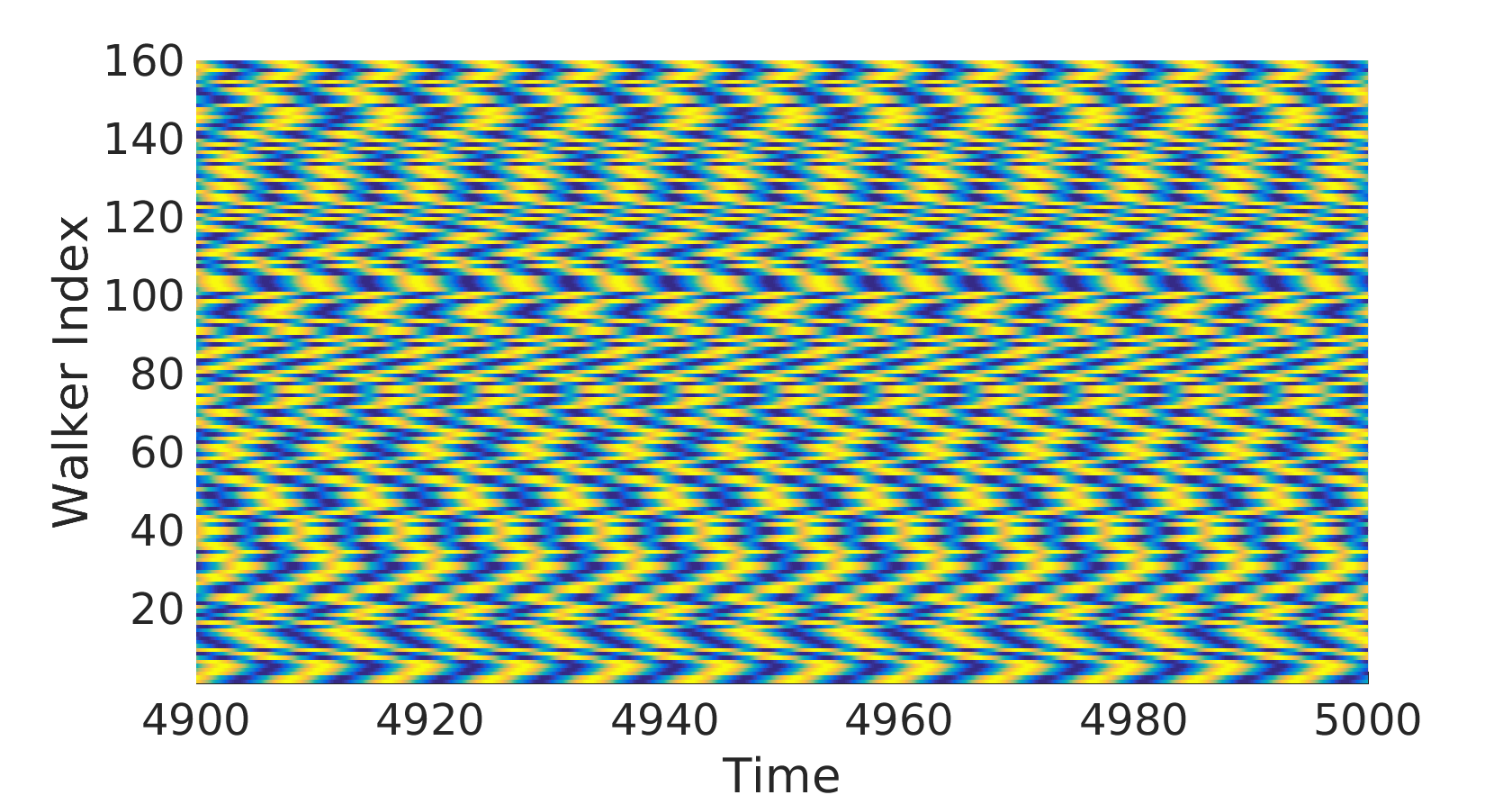} \includegraphics[width=0.49\textwidth]{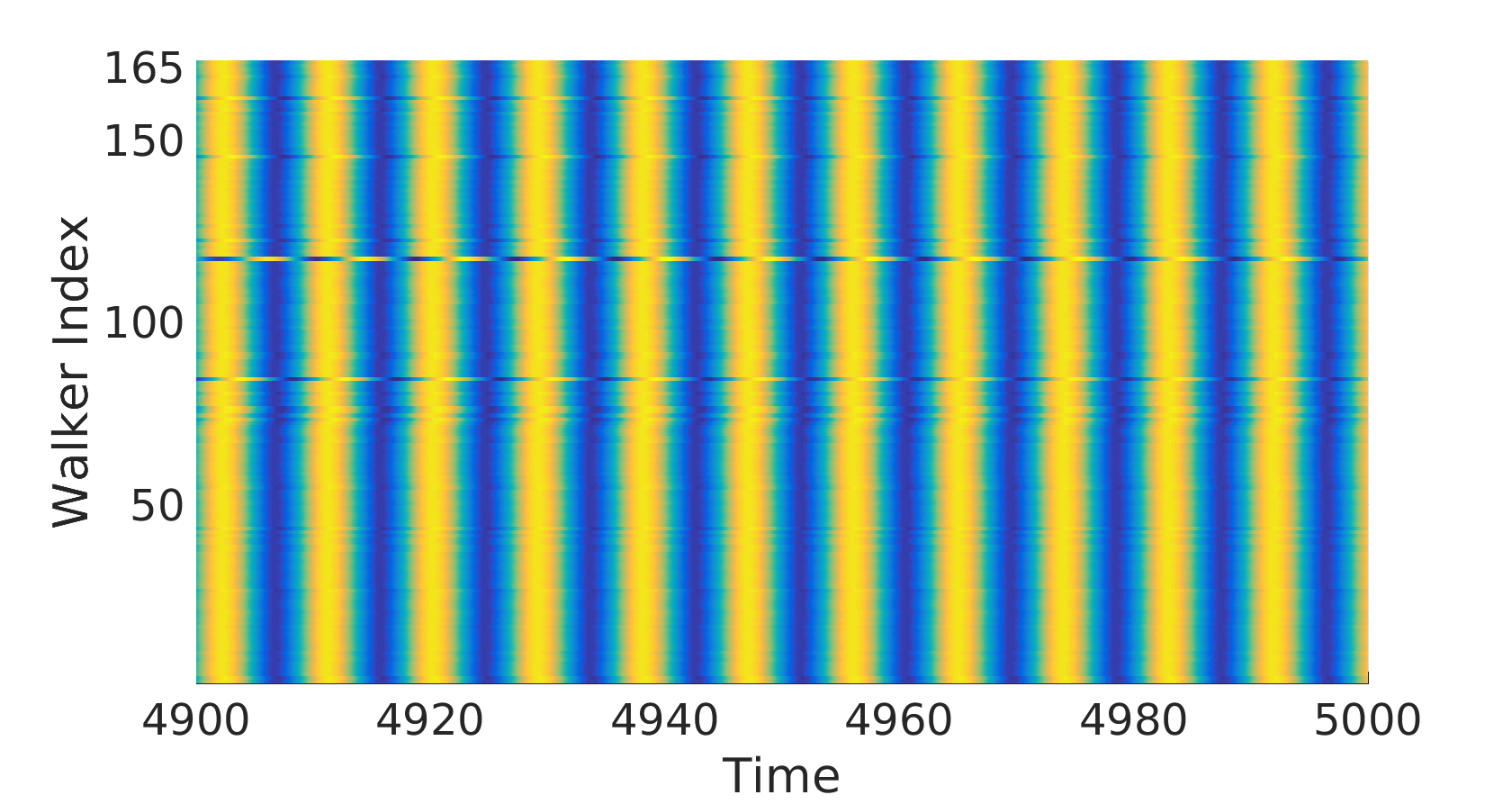}
\end{center}
     \caption{Pedestrian motions represented by an amplitude color plot for a crowd size before ($n=160$) (left) and after $n=166$ (right) the onset of phase locking at the critical number of walkers ($n\approx165$). The full video demonstrating the onset of phase locking as a function of the addition of pedestrians on the bridge for identical models (\ref{eq:b1}) is given in the Supplement  \cite{video-van-der-pol}.} \label{fig4}
       \end{figure}
\begin{figure}
	\begin{center}
\includegraphics[width=0.49\textwidth]{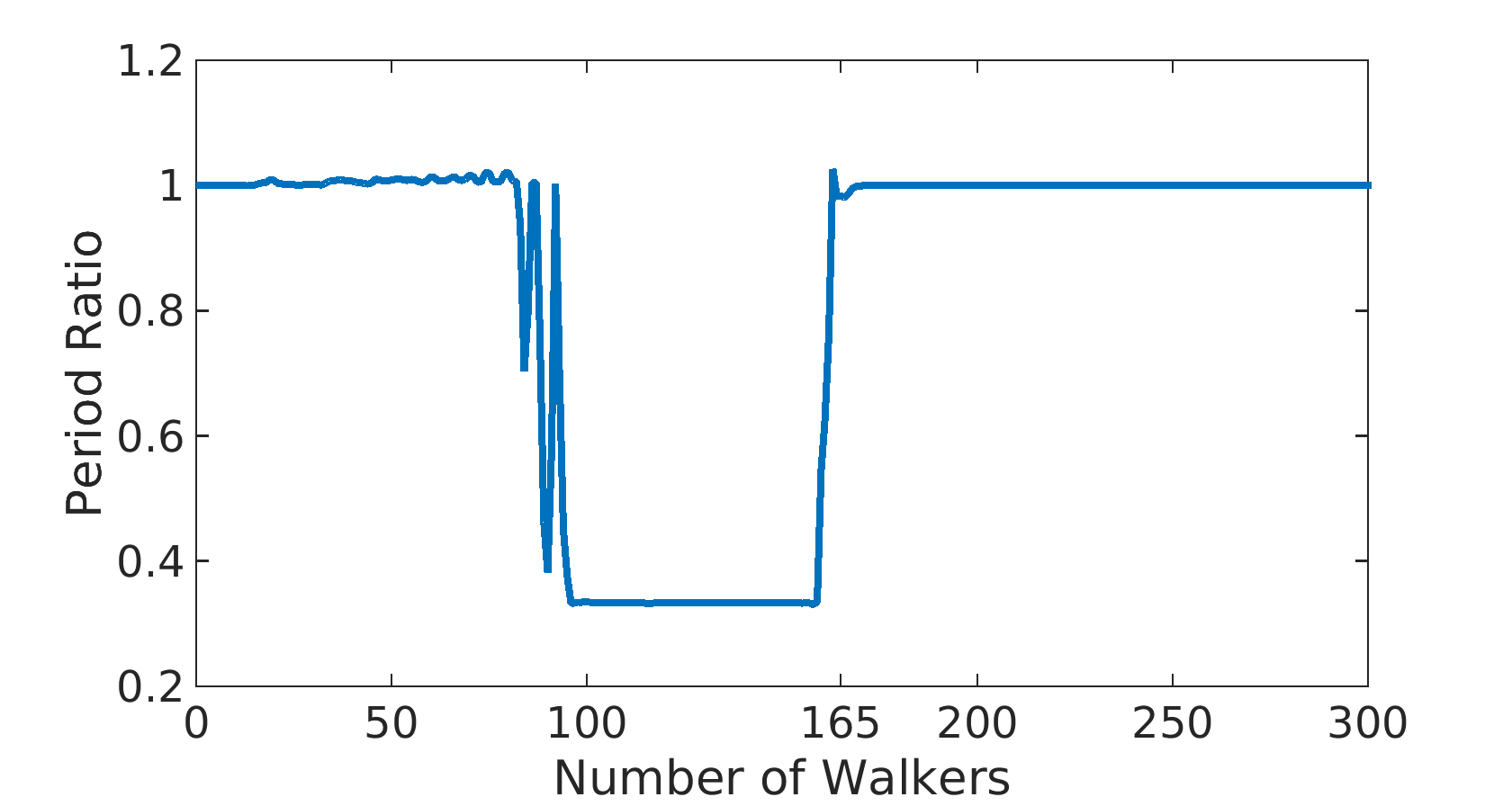} \includegraphics[width=0.49\textwidth]{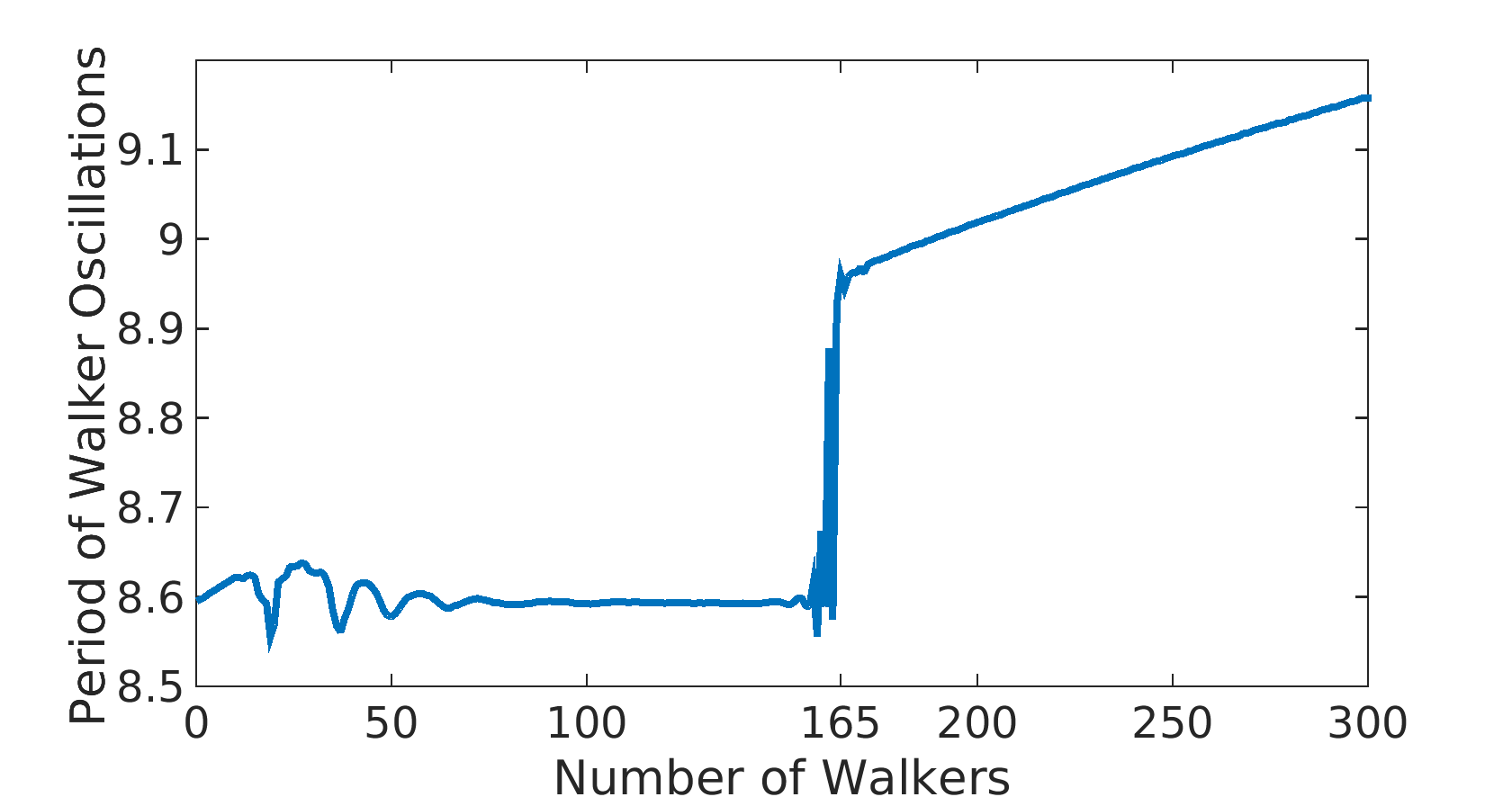}
\end{center}
     \caption{(Left).  Ratio of periods of bridge and pedestrian oscillations.  When phase-locked at $n = 165$, the pedestrians begin walking in phase with the oscillations of the bridge (indicated by a period ratio of $1$). (Right). Effect of bridge feedback on the period of walker oscillations. As the total mass of the pedestrian-bridge system increases with $n$, the period of phase-locked oscillations also increases. }\label{fig5}
       \end{figure}
The above analysis provides a clear example of the threshold effect when the increasing crowd size induces the abrupt onset of bridge wobbling. This onset is caused by the emergence of phase-locking among the pedestrians and the bridge. According to our analysis and numerics, this phase-locking occurs at frequency one if the intrinsic parameters of the pedestrian motion and the bridge are chosen in accordance with the balance condition (\ref{eq:i12}).

A natural question which arises in this context is whether or not the pedestrian-bridge system (\ref{eq:b1}) still exhibits the threshold effect and the emergence of phase-locking if the balance condition cannot be met. In the following, we will give a positive answer to this question and demonstrate that even though our analytical conditions are no longer applicable, the pedestrian-bridge system can still be tuned to exhibit the very same effect of phase locking. However, this phase locking occurs at a different frequency which can only be assessed numerically.  This does not come as a surprise, given the complexity of the nonlinear system (\ref{eq:b1}) with arbitrarily chosen parameters.

We knowingly consider the worst case scenario and set the natural bridge frequency $\Omega=1$ at which the balance condition (\ref{eq:i12}) has a singularity and is undefined. As a result, the analytical conditions (\ref{eq:i10})-(\ref{eq:i12}) cannot be satisfied. Yet, the pedestrian-bridge system (\ref{eq:b1}) can be tuned to exhibit phase-locked oscillations and therefore induce bridge wobbling at the desired critical number of pedestrians $n^*=165,$ however, at a frequency different from one. Similarly to Fig.~3, Fig.~\ref{fig3} (top) indicates this sudden wobbling in the bridge after the number of walkers exceeds the threshold ($n=165$). In addition to the threshold effect of adding pedestrians to the bridge, Fig. \ref{fig3} (top) shows a minor increase in bridge wobbling oscillations for $n\in[10,40]$.  This increase is caused by a cluster of phase-locked walkers paired with a cluster of unsynchronized walkers.  In dynamical systems literature, this pairing of a synchronized cluster with an asynchronous cluster (despite the homogeneity of oscillators) is often called a ``chimera'' state \cite{chimera-1}. Notice, that this cluster of phase-locked pedestrians disintegrates with a further increased crowd size and stops this low-amplitude wobbling, until the critical crowd size $n_c$ is reached.
To give more details on the onset of pedestrian phase-locking, Fig.~\ref{fig3} (middle) displays the evolution of the average phase mismatch, $\Delta \Phi = \frac{\sum \limits_{i \neq j} |\varphi_i - \varphi_j|}{n(n-1)},$ where $\varphi_i$ is the normalized period of oscillations for walker $x_i.$ Fig.~\ref{fig3} (bottom) presents the largest transversal Lyapunov exponent $\lambda_\perp$ for the stability of complete synchronization among the pedestrians as a function of the crowd size. Notice that the pedestrians become completely phase-locked (synchronized) only when the crowd size reaches its critical size $n_c$ and the Lyapunov exponent becomes negative.
Figure \ref{fig4} shows the transition from the asynchronous pedestrian motion to fully phase-locked pedestrian behavior for crowd sizes right below and above the threshold value.
While phase-locking occurs at a frequency different from one, the ratio of the periods of oscillation for the bridge and the pedestrians becomes one (see Fig. \ref{fig5} (left)). That is,
when the pedestrians synchronize, they begin walking in phase with the oscillations of the bridge
(hence the ratio of the periods of the bridge and walker oscillations is $1$). Figure \ref{fig5} (right) also shows that after the onset of phase-locking, the addition of more pedestrians causes the period of oscillations for both the pedestrians and the bridge to increase as the pedestrian-bridge system becomes heavier. This generates the increase in the amplitude of bridge oscillations shown in Fig. \ref{fig3}.

\subsubsection{Non-identical Van der Pol-type walkers}
Our theory can also be extended to the pedestrian-bridge system (\ref{eq:b1}) with non-identical pedestrians who have different natural frequencies $\omega_i,$ randomly distributed within the
range $[\omega_{-},\omega_{+}].$ Following the steps of the above study, we seek to determine the conditions that yield phase-locked solutions $x_i(t)=B_i\sin(t+\varphi_i),$ $i=1,...,n$ and $y(t)=A\sin(t+\psi)$
and reveal the threshold effect which induces bridge wobbling at a sufficiently large crowd size.

Substituting these solutions for $x_i$ and $y$ into the system (\ref{eq:b1}), we obtain the balance conditions, similar to those from the identical pedestrian case:
\begin{equation}
A^2=B_i^2(\omega_i^2-1)^2+\lambda^2(B_i^2-a^2)^2 B_i^2,
\label{eq:n1}
\end{equation}
\begin{equation}
\tan(\varphi_i-\psi)=-\lambda\frac{B_i^2-a^2}{\omega_i^2-1}.
\label{eq:n2}
\end{equation}
\begin{equation}
\begin{array}{l}
A^2=\frac{(rn)^2}{\Delta}\bar{B}^2,
\end{array}
\label{eq:n3}
\end{equation}
\begin{equation}
\begin{array}{l}
\tan(\bar{\varphi}-\psi)=\frac{2h}{1-\Omega^2},
\end{array}
\label{eq:n4}
\end{equation}
where $\bar{B}=\frac{1}{n} \sum \limits_{i=1}^n B_{i}$ and $\bar{\varphi}=\frac{1}{n} \sum \limits_{i=1}^n \varphi_{i}$ are the mean amplitude and phase of the group of pedestrians, respectively.
 \begin{figure*}[ht]
	\begin{center}
\includegraphics[width=0.65\textwidth]{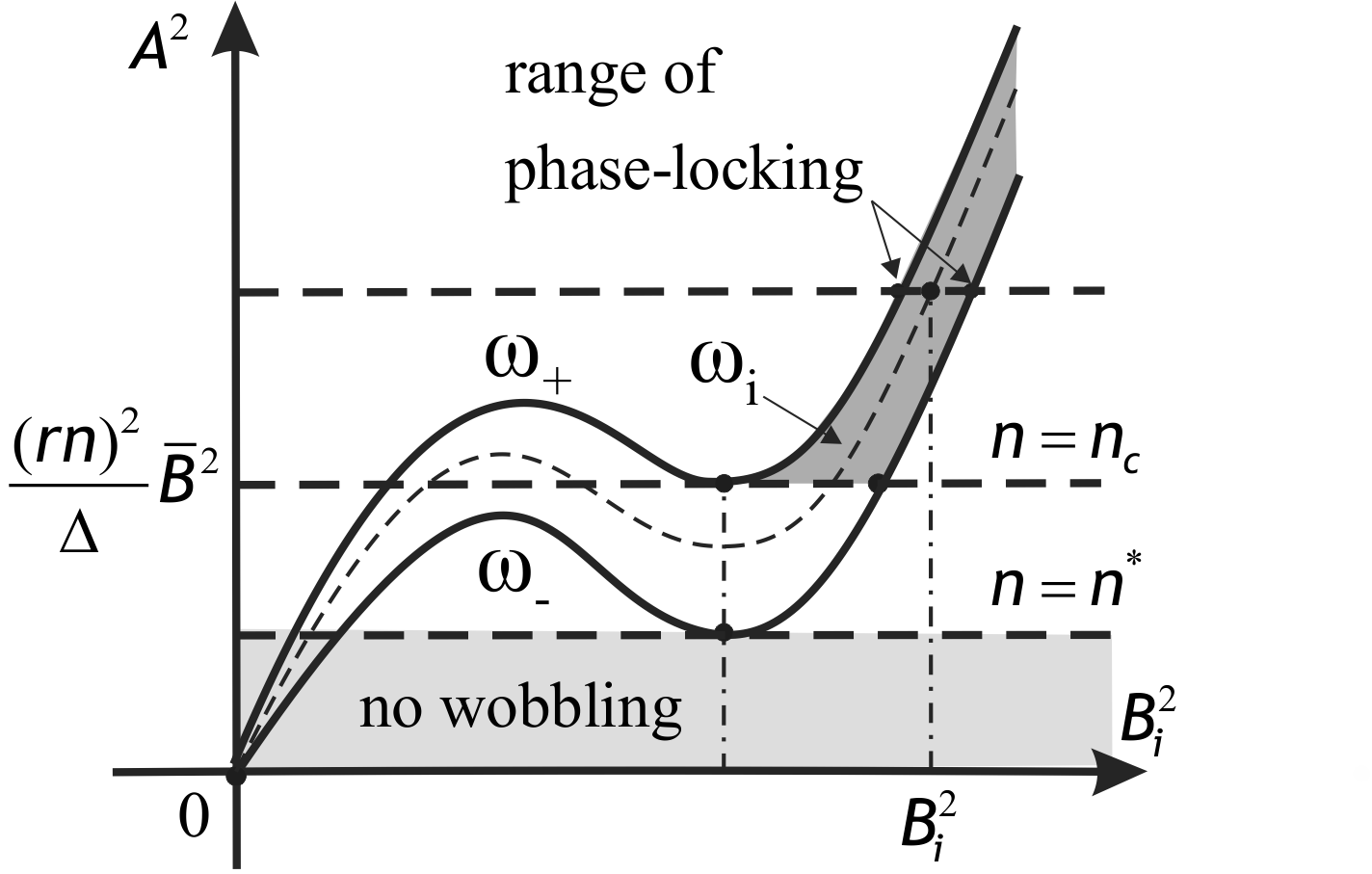}
\end{center}
     \caption{Nonidentical oscillators with $\omega_i \in [\omega_{-},\omega_{+}].$ Schematic diagram similar to the identical case of Fig.~2, where the cubic balance curve and inclined lines become a cubic strip and horizontal lines, respectively. The dark gray area: the range of phase-locking, expressed through permissible amplitudes $B_i$ and phases via (\ref{eq:n5})-(\ref{eq:n6}). } \label{non}
     \end{figure*}
     Our goal is to identify the amplitudes $B_i$ and $\varphi_i$ which are consistent with the amplitude and phase conditions (\ref{eq:n1})-(\ref{eq:n2}) and at the same time yield
the desired $\bar{B}$ and $\bar{\varphi}$ to satisfy the conditions (\ref{eq:n3})-(\ref{eq:n4}). Toward this goal, we compare the amplitude conditions  (\ref{eq:n1})-(\ref{eq:n3})
\begin{equation}
B_i^2(\omega_i^2-1)^2+\lambda^2(B_i^2-a^2)^2 B_i^2=\frac{(rn)^2}{\Delta}\bar{B}^2.
\label{eq:n5}
\end{equation}
Solving this equation graphically (Fig.~\ref{non}), we obtain a cubic strip of the balance curves $\omega_i$ which represent the LHS of (\ref{eq:n5}) and a horizontal balance line defined by the RHS of (\ref{eq:n5}). The intersection point between the balance curve $\omega_i$ with the horizontal line yields the amplitude $B_i$ for the phase-locked oscillation $x_i(t).$ Similarly to the identical pedestrian case, Fig.~\ref{non} reveals the bifurcation mechanism for the abrupt onset of the bridge wobbling, defined by phase-locking among all $n$ oscillators that appears when $n$ exceeds the critical number $n_c$ (when the horizontal balance line (\ref{eq:n3}) intersects the upper curve $\omega_+$). Note that the intersections with the left and middle branches of the cubic line $\omega_+$ also yield permissible solutions; however, they are unstable and a saddle, respectively. When the horizontal balance line (\ref{eq:b3}) intersects the cubic stripe at a level between $n^*$ and $n_c$, only a fraction of balance curves with $\omega_i$ lie above this line such that only a group of pedestrians is phase-locked and only small-amplitude wobbling can appear from this partially phase-locked state.

A potential problem in solving the condition  (\ref{eq:n5}) for $B^2_i$ is that the mean amplitude $\bar{B}$ is a function of $B_i,$ $i=1,..,n.$ Therefore, the values $B_1,B_2,...,B_n$ which are obtained from the intersections of the corresponding curves, do not necessarily sum up to match $\bar{B}.$ To get around this problem, we use the following trick. Given the values $B_1,B_2,...,B_n,$ we calculate its exact mean amplitude $\bar{B}.$ This value
differs from the original value of $\bar{B}$ which was initially chosen to set the level of the horizontal line $A^2=\frac{(rn)^2}{\Delta}\bar{B}^2$ which generated $B_1,B_2,...,B_n.$ To resolve this issue,
we can always adjust the parameter $\Delta,$ by varying $\Omega$ or $h$ to preserve the same level of the horizontal line $A^2=\frac{(rn)^2}{\Delta}\bar{B}^2,$ with the updated mean amplitude $\bar{B}$ which matches the values of $B_1,B_2,...,B_n.$ Notice that this change of $\Delta$ does not affect the location of the cubic curves such that the values of $B_1,B_2,...,B_n$ remain intact. As a result of this procedure, we obtain the values $B_i$ whose the mean amplitude matches the condition (\ref{eq:n5}) and yields the bridge amplitude $A.$ Therefore, it remains to satisfy the phase balances (\ref{eq:n2})-(\ref{eq:n4}) to identify the phases $\varphi_i,$ $i=1,..,n$ and $\psi.$

Comparing the phase balance conditions (\ref{eq:n2}) and (\ref{eq:n4}), we derive the following equations
\begin{equation}
\varphi_i+\bar{\varphi}=b_i,\;i=1,...,n,
\label{eq:n6}
\end{equation}
where $b_i=\arctan \lambda\frac{B_i^2-a^2}{1-\omega_i^2}-\arctan \frac{2h}{\Omega^2-1}$ are known constants, given the fixed values of $B_i$ and parameters $\lambda,\omega_i,h,$ and $\Omega.$ As $\bar{\varphi}=\frac{1}{n}(\varphi_1+...+\varphi_n),$ finding $\varphi_i,$ $i=1,...,n$ amounts to solving the system of $n$ linear equations (\ref{eq:n6}). Therefore, the phases $\varphi_i$ can be calculated even for a large crowd size $n$ with only moderate effort. This will also allow us to calculate the mean phase $\bar{\varphi}$ and find the unknown bridge phase $\psi$ via (\ref{eq:n4}):
\begin{equation}
\begin{array}{l}
\psi=\bar{\varphi}-\arctan \frac{2h}{1-\Omega^2}.
\end{array}
\label{eq:n7}
\end{equation}
The above procedure suggests a way of calculating the values of $B_i,$ $\varphi_i,$ $A,$ $\psi$ and choosing the parameters of the pedestrian-bridge system (\ref{eq:b1}) which yield the phase-locked solutions
$x_i(t)$ and $y(t)$ with frequency one and phases $\varphi_i$ which emerge when the crowd size reaches the critical value $n_c.$  In this way, the distribution of natural frequencies $\omega_i=[\omega_{-},\omega_{+}]$ translates
into a distribution of phases $\varphi_i$ which preserve the phase balances and the phase-locked solution with frequency one, similar to the identical pedestrian case.
While this procedure is direct, it gives implicit conditions for the existence and emergence of the phase-locked solutions and their dependence on the parameters of the pedestrian-bridge system. However, necessary conditions for the absence of the phase-locked solutions $x_i(t)$ and $y(t)$ can be given explicitly as follows.

Notice that the mean amplitude is $\bar{B}=\frac{1}{n}(B_{1}+B_{2}+...+B_{n})$ such that its square can be bounded by means of the Cauchy-Schwarz inequality:
$\bar{B}^2=\frac{1}{n^2}(B_{1}+B_{2}+...+B_{n})^2 \le \frac{1}{n}(B_{1}^2+B_{2}^2+...+B_{n}^2).$ Therefore, the balance equation (\ref{eq:n3}) can be turned into the inequality:
\begin{equation}
A^2 \le \frac{(rn)^2}{\Delta}\left [  \frac{1}{n}(B_{1}^2+B_{2}^2+...+B_{n}^2) \right]=\frac{(rn)^2}{\Delta} \bar{B^2}.
\label{eq:n8}
\end{equation}
An elegant way to obtain the bounds that guarantee the absence of phase-locked solutions is to combine the amplitude balance equation (\ref{eq:n1}) and the inequality (\ref{eq:n8})
\begin{equation}
\begin{array}{ll}
\frac{(rn)^2}{\Delta}\left [  \frac{1}{n}(B_{1}^2+B_{2}^2+...+B_{n}^2) \right]
\le &B_i^2(\omega_i^2-1)^2+\\
&\lambda(B_i^2-a^2)^2B_i^2.
\end{array}
\label{eq:n9}
\end{equation}

We use the lowest frequency $\omega_-$ from the interval $\omega_i\in[\omega_-,\,\omega_+]$ and strengthen the inequality by replacing $B_i^2$ with the mean $\bar{B^2}$ on the RHS to obtain
the lower bound curve for all $B_i$: $\frac{(rn)^2}{\Delta} \bar{B^2} \le \bar{B^2}(\omega_{-}^2-1)^2+\lambda(\bar{B^2}-a^2)\bar{B}^2.$ Consequently, we get the condition
$\frac{(rn)^2}{\Delta} -(\omega_{-}^2-1)^2 \le \lambda(\bar{B^2}-a^2).$ Notice that if the LHS of this inequality is less than 0, the inequality holds true for any amplitude $\bar{B^2}$. Therefore, we get the bound
$\frac{(rn)^2}{\Delta} -(\omega_{-}^2-1)^2 < 0$ which is similar to that from the identical pedestrian case (\ref{eq:i10}) where the frequency $\omega$ is replaced with the lowest frequency $\omega_{-}:$
\begin{equation}
\frac{mn^*}{mn^*+M} <|\omega_{-}^2-1|\sqrt{(\Omega^2-1)^2+4h^2},
\label{eq:b6}
\end{equation}
where $n^*$ is the bound on the number of pedestrians below which the phase-locked solutions are absent and the wobbling of the bridge is practically zero (cf. Fig.~\ref{non}). Explicit in the parameters of the bridge, this condition can be used as a safety guideline for designing pedestrian bridges or limiting the maximum occupancy of an existing bridge. It also suggests that pedestrians with lower frequencies of their lateral gait are more prone to locking-in with the bridge motion.
   \begin{figure}
       	\begin{center}
\includegraphics[width=0.49\textwidth]{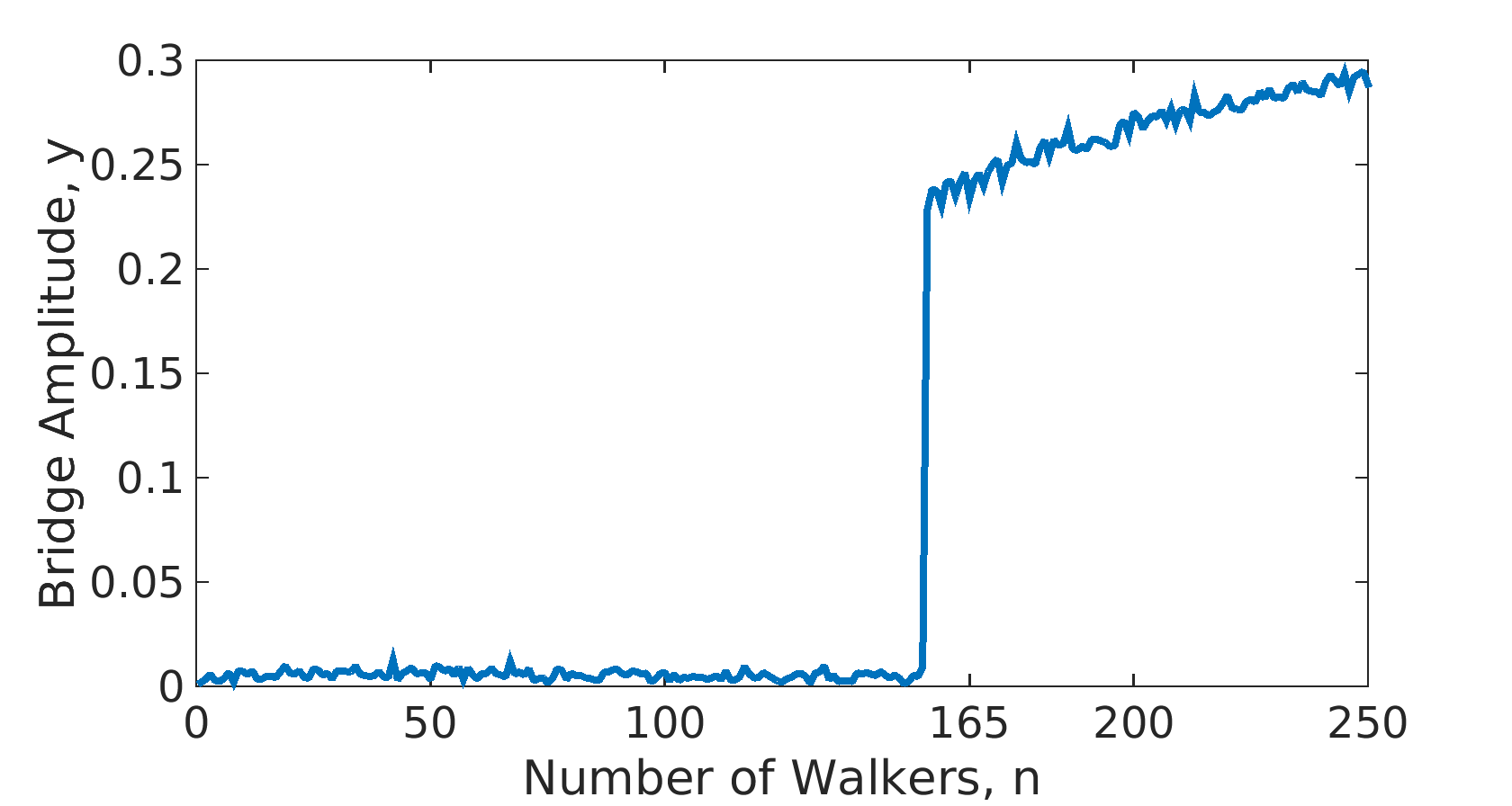} \includegraphics[width=0.49\textwidth]{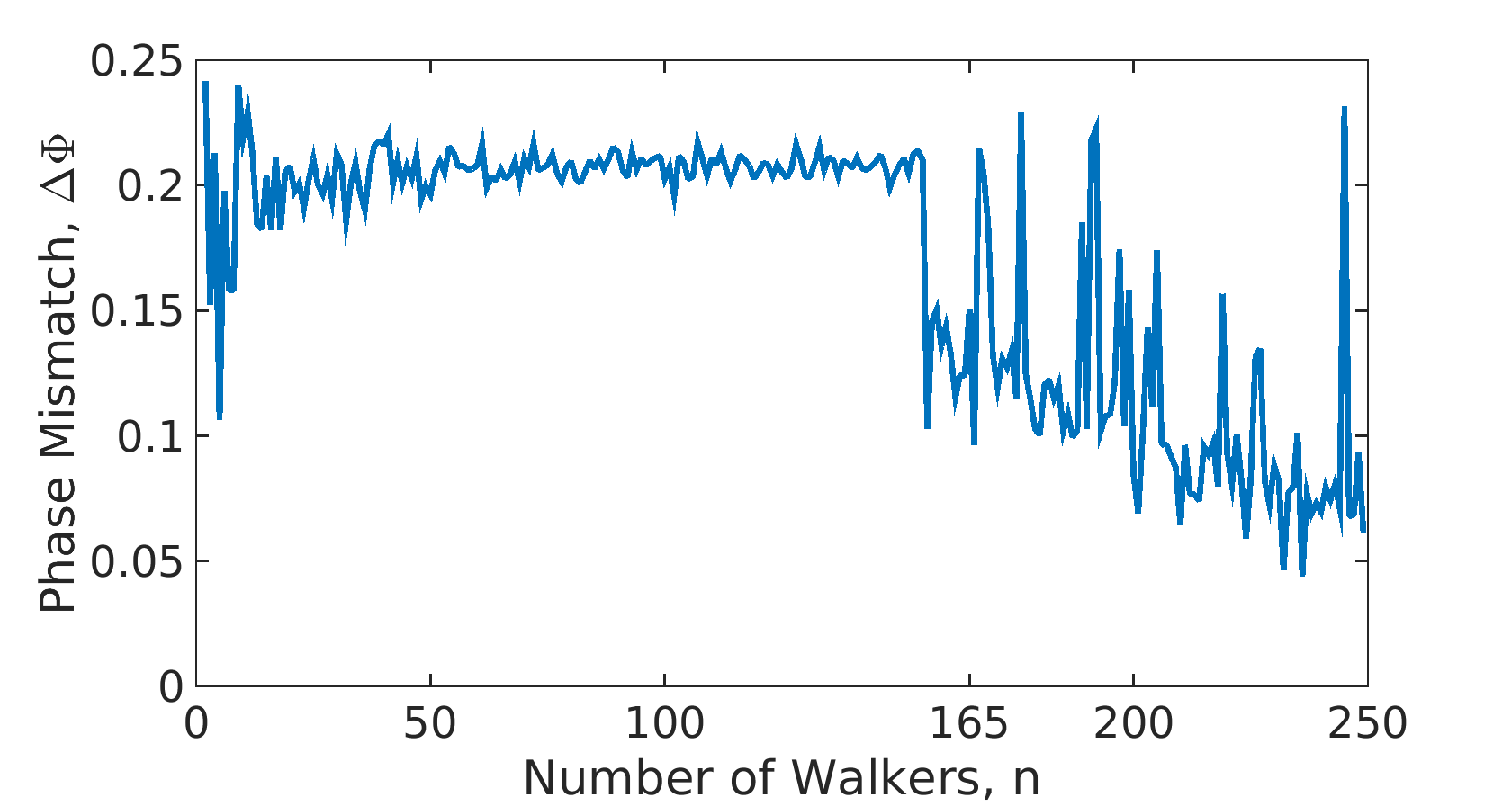}
\end{center}
     \caption{Inverted pendulum model (\ref{eq:1})-(\ref{eq:3}) of non-identical pedestrians with randomly chosen frequencies $\omega_i\in [0.6935,0.7665]$ ($10\%$ mismatch). Diagrams similar to
      Fig. \ref{fig3}.  The onset of bridge oscillations is accompanied by a drop in the average phase difference between the pedestrian's foot adjustment. The initial drop corresponding to the initiation of the bridge wobbling and partial phase-locking is less significant, compared to the well-established phase-locking at larger crowd sizes over $200$ pedestrians.
     The individual pedestrian parameters are $\lambda=2.8,$ $p=1$, $a=1.$ Other parameters are as in Fig. \ref{fig3}.
     Model (\ref{eq:1})-(\ref{eq:3}) with identical walkers produce similar curves with nearly the same critical crowd size, however the phase difference drops to zero (as complete phase-locking is possible for identical oscillators) [not shown].}
       \label{fig6}
        \end{figure}

\section{Non-identical inverted pendulum walkers}\label{section 4}
\begin{figure}
       	\begin{center}
       	\includegraphics[width=0.49\textwidth]{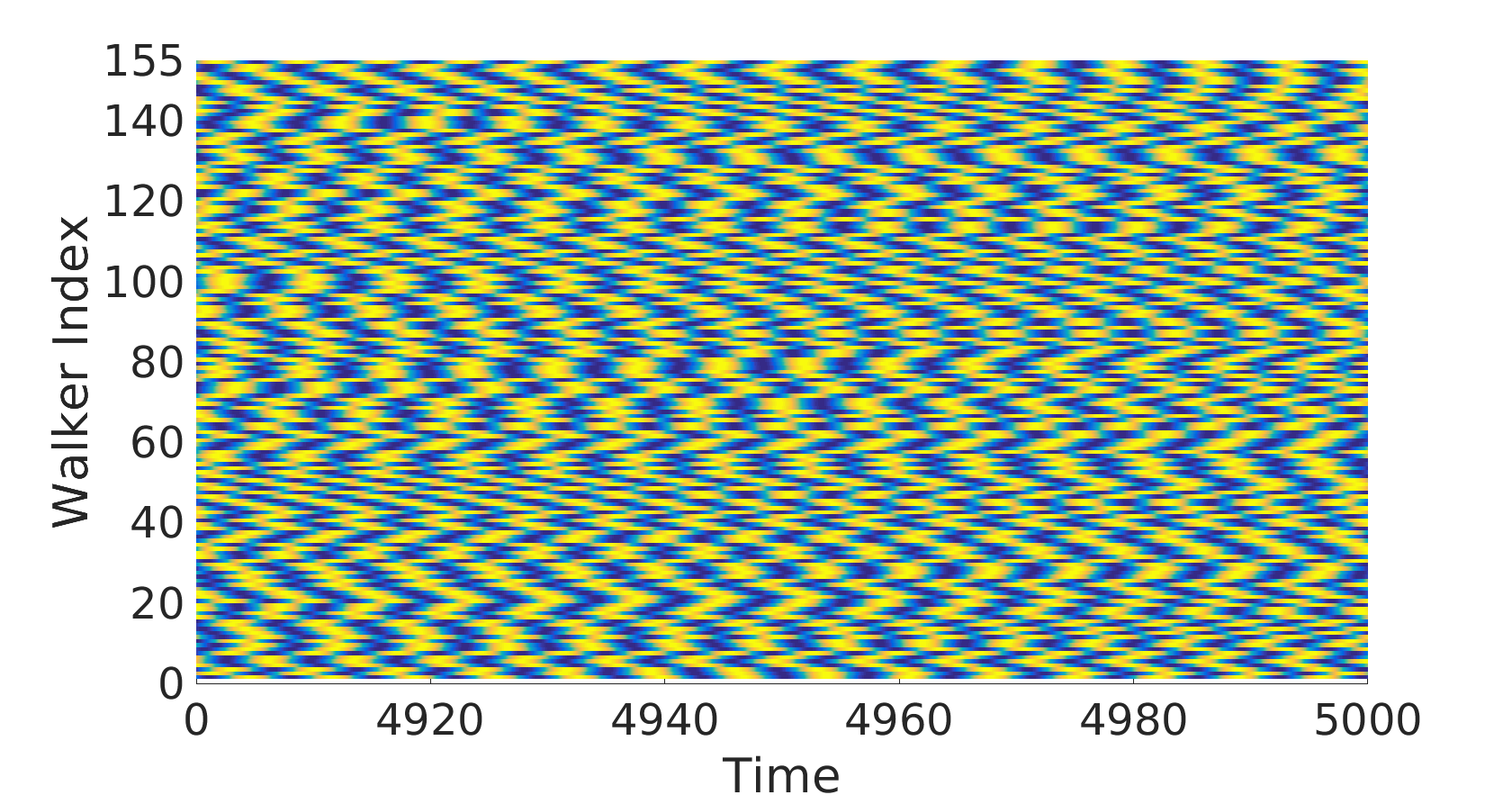} \includegraphics[width=0.49\textwidth]{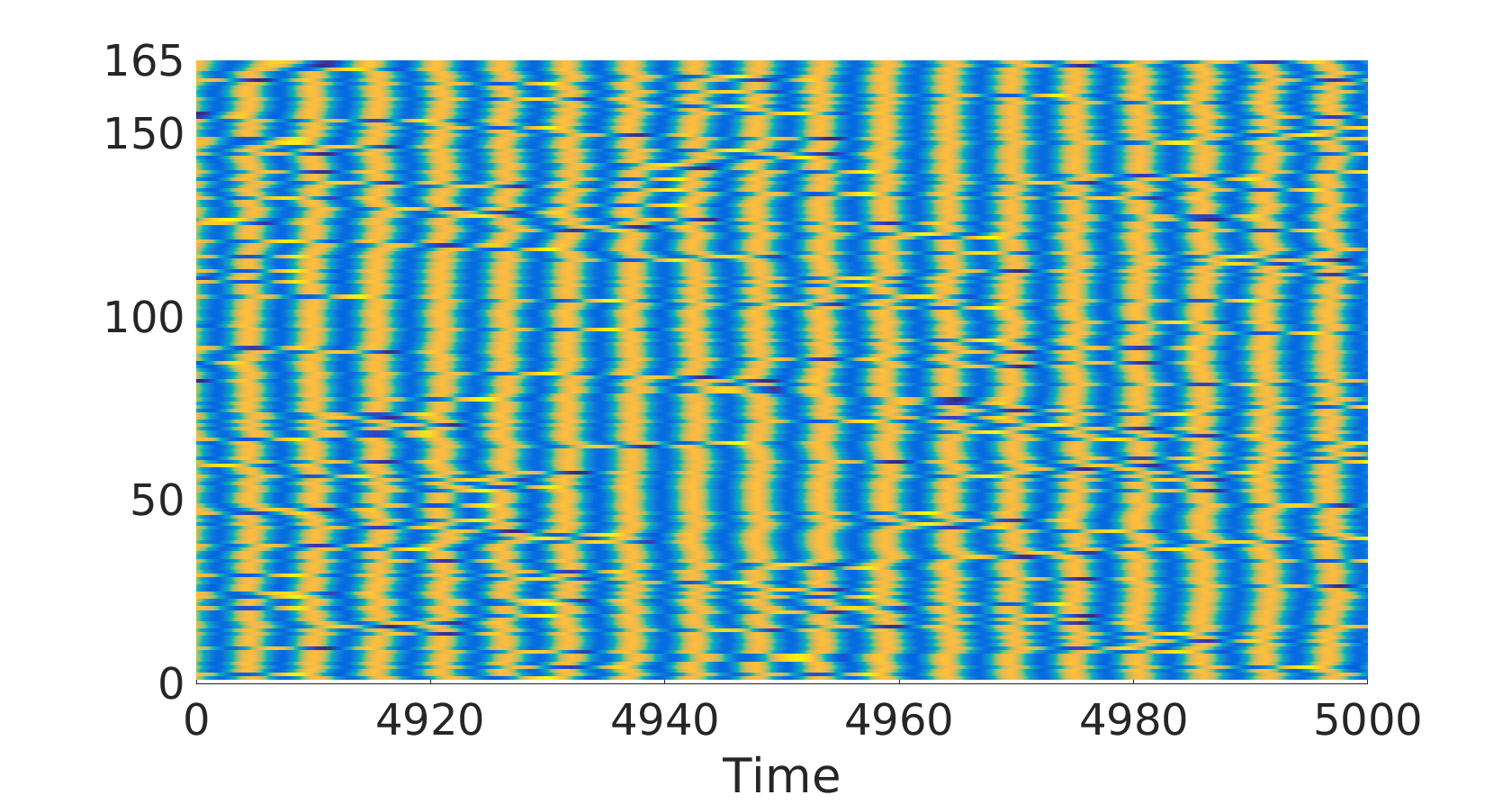}
       	\end{center}
       	\caption{Diagrams for the mismatched inverted pendulum pedestrian-bridge model (\ref{eq:1})-(\ref{eq:3}), similar to Fig.~ \ref{fig4}. The full videos for both identical and mismatched inverted pendulum models are given in the Supplement \cite{video-inverted}.} \label{fig7}
       \end{figure}

We return to the most realistic case of bio-inspired inverted pendulum pedestrian-bridge model (\ref{eq:1})-(\ref{eq:3}) with a $10\%$ frequency mismatch (see Fig.~\ref{fig6}). We randomly choose the frequencies from
the interval $\omega_i\in [0.6935,0.7665]$ which is centered around the frequency $\omega=0.73$ used in the above numerical simulations of the identical  Van der Pol-type pedestrian models (\ref{eq:b1}) (cf. Fig.~4). We choose other free individual pedestrian parameters as follows $\lambda_i=\lambda=2.8,$ $p=2$, $a=1.$ This choice together with the fixed parameters for the London Millennium Bridge (cf. Fig. \ref{fig3} and Fig. \ref{fig6}) gives the threshold value around $n_c=160$. Therefore, we consistently obtain nearly the same threshold for both Van der Pol and  inverted pendulum models. Figures \ref{fig6} and \ref{fig7} indicate that well-developed phase-locking is inevitably present when the number of pedestrians exceeds $n_c$; however, the abrupt initiation of bridge wobbling right at the edge of instability around the critical number does not exactly coincide with total phase-locking among the pedestrians and is accompanied by multiple phase slips and detuning. This suggests that pedestrian phase-locking is crucial for a highly inertial system such as the London Millennium bridge weighing over 113 tons to develop significant wobbling. However, its initiation mechanism at the edge of instability can be more complicated than simple phase-locking. In particular, the balance control based on the lateral position of foot placement can initiate low-amplitude bridge wobbling, prior to the onset of crowd synchrony at larger crowd sizes as suggested in \cite{macdonald-inverted}.

\section{Discussions and Conclusions}
\label{section 5}
The history of pedestrian and suspension bridges is full of dramatic events. The most recent examples are
(i) pedestrian induced vibrations during the openings of the Solf\'erino Bridge in Paris in 1999 \cite{danbon} and the London Millennium Bridge in 2000, and the increased swaying of the Squibb Park Bridge in Brooklyn in 2014 \cite{brooklyn}, and (ii) the Volga Bridge \cite{world} in the Russian city of Volgograd (formerly, Stalingrad) (cost $\$$396 million and $13$ years to build) which experienced wind-induced resonance vibrations in 2011 soon after its opening and was shutdown for expensive repairs. Parallels between wind and crowd loading of bridges have been widely discussed \cite{robie} as intensive research on the origin of resonant vibrations, caused by the pedestrian-bridge interaction and wind-induced oscillations can have an enormous safety and ecomonic impact.

In this paper, we have contributed towards understanding the dynamics of pedestrian locomotion and its interaction with the bridge structure. We have proposed two models where the individual pedestrian is represented by
a bio-mechanically inspired inverted pendulum model (\ref{eq:3}) and its simplified Van der Pol analog. We have used the parameters close to those of the London Millennium Bridge to verify the popular explanation that the wobbling of London Millennium Bridge was initiated by phase synchronization of pedestrians falling into step with the bridge's oscillations. The analysis of both models has indicated the importance of the inclusion of foot force impacts for a more accurate prediction of the threshold effect. Surprisingly, the pedestrian-bridge system with a Van der Pol-type oscillator as an individual pedestrian model allows for the rigorous analytical analysis of phase-locked solutions, even though periodic solutions in the individual Van der Pol-type oscillator cannot be found in closed form.

Our numerical study of the inverted pendulum pedestrian-bridge model confirms that crowd phase-locking was necessary for the London Millennium Bridge to wobble significantly, especially at intermediate and large amplitudes.
However, our simulations indicate that the initiation of wobbling is accompanied by tuning and detuning of pedestrians footstep such that phase lock-in mechanism might not necessarily be the main cause of initial small-amplitude wobbling. A rigorous analysis of the inverted pendulum-bridge model to reveal secondary harmonics at which the wobbling can be initiated without total phase-locking is a subject of future study.

The initiation of wobbling without crowd phase locking was previously observed during periods of instability of the Singapore Airport's Changi Mezzanine Bridge \cite{danbon} and the Clifton Suspension Bridge {\cite{clifton}.
Both bridges experienced crowd-induced vibrations at a bridge frequency different from the averaged frequency of the pedestrians, while the pedestrians continued to walk without visible phase-locking  {\cite{macdonald-inverted}.
Our recent results \cite{belykh2016bistable} on the ability of a single pedestrian to initiate bridge wobbling when switching from one gait to another may shed light on the initiation of small wobbling without crowd synchrony. More detailed models of pedestrian-bridge interaction, that incorporate additional factors such as, for example, an extra degree of freedom that accounts for the knee control in the individual pedestrian model and person-to-person visual communication and pace slowing in dense crowds, may also suggest an additional insight into the origin of this small-amplitude wobbling.  From the dynamical systems perspective, this is an important piecewise smooth problem that requires careful mathematical study of the dynamics of non-smooth oscillators (pedestrians), with the nonlinear coupling within the crowd and a bi-direction interaction with a bridge structure.

Our theory and models should help engineers to better understand the dynamical impact of crowd collective behavior and guarantee the comfort level of pedestrians on a bridge. This requirement represents a major challenge as the natural frequency of an aesthetically pleasing lively bridge often falls into a critical frequency range of pedestrian phase-locking. These frequencies cannot be identified through the conventional linear calculations that might lead to faulty designs.

\section{Acknowledgment}
We gratefully acknowledge guidance with existing bio-inspired inverted pendulum models of pedestrian gait
and the role of balance control from Alan Champneys and John Macdonald, both of the University of Bristol, UK.
We thank Dario Feliciangeli, a bridge engineer with Mott Macdonald Group, for pointing us to the US Guide Specifications for the design of pedestrian bridges and useful discussions.

\bibliographystyle{siamplain}

\bibliography{references}

\end{document}